\pdfoutput=1
\documentclass[12pt,letter]{article}

\usepackage{ifthen} 
\newboolean{pdflatex}
\setboolean{pdflatex}{true} 

\newboolean{articletitles}
\setboolean{articletitles}{true} 

\newboolean{uprightparticles}
\setboolean{uprightparticles}{false} 

\def\paperauthors{R.F.Lebed, T.Skwarnicki}
\def\paperasciititle{Summary of Topical Group on Hadron Spectroscopy (RF07) Rare Processes and Precision Frontier of Snowmass 2021} 
\def\papertitle{Summary of Topical Group\\ on Hadron Spectroscopy (RF07) \\Rare Processes and Precision Frontier\\ of Snowmass 2021} 
\def\paperkeywords{{High Energy Physics}, {Hadron Spectroscopy}} 
\def\papercopyright{\quad } 
\def\paperlicence{CC BY 4.0 licence}  
\def\paperlicenceurl{https://creativecommons.org/licenses/by/4.0/}

\usepackage[top=1in, bottom=1.25in, left=1in, right=1in]{geometry}

\usepackage{microtype}
\usepackage{lineno}  
\usepackage{xspace} 
\usepackage{caption} 

\usepackage{graphicx}  
\usepackage{color}
\usepackage{colortbl}
\graphicspath{{./figs/}} 

\usepackage{amsmath} 
\usepackage{amssymb}
\usepackage{amsfonts}
\usepackage{upgreek} 

\newcommand*\patchAmsMathEnvironmentForLineno[1]{%
\expandafter\let\csname old#1\expandafter\endcsname\csname #1\endcsname
\expandafter\let\csname oldend#1\expandafter\endcsname\csname
end#1\endcsname
 \renewenvironment{#1}%
   {\linenomath\csname old#1\endcsname}%
   {\csname oldend#1\endcsname\endlinenomath}%
}
\newcommand*\patchBothAmsMathEnvironmentsForLineno[1]{%
  \patchAmsMathEnvironmentForLineno{#1}%
  \patchAmsMathEnvironmentForLineno{#1*}%
}
\AtBeginDocument{%
\patchBothAmsMathEnvironmentsForLineno{equation}%
\patchBothAmsMathEnvironmentsForLineno{align}%
\patchBothAmsMathEnvironmentsForLineno{flalign}%
\patchBothAmsMathEnvironmentsForLineno{alignat}%
\patchBothAmsMathEnvironmentsForLineno{gather}%
\patchBothAmsMathEnvironmentsForLineno{multline}%
\patchBothAmsMathEnvironmentsForLineno{eqnarray}%
}

\usepackage{hyperxmp}

\usepackage[pdftex,
            pdfauthor={\paperauthors},
            pdftitle={\paperasciititle},
            pdfkeywords={\paperkeywords},
            pdfcopyright={Copyright (C) \papercopyright},
            pdflicenseurl={\paperlicenceurl}]{hyperref}

\usepackage[colorinlistoftodos,textsize=scriptsize]{todonotes}

\usepackage[bottom,flushmargin,hang,multiple]{footmisc}

\usepackage[all]{hypcap} 

\usepackage{cite} 
\usepackage{mciteplus}
\usepackage{longtable} 

\begin{document}

\renewcommand{\thefootnote}{\fnsymbol{footnote}}
\setcounter{footnote}{1}


\begin{titlepage}
\pagenumbering{roman}

\vspace*{-1.5cm}
\centerline{\large Snowmass 2021 (DPF Community Planning Exercise) }
\vspace*{0.5cm}
\noindent
\begin{tabular*}{\linewidth}{lc@{\extracolsep{\fill}}r@{\extracolsep{0pt}}}
\vspace*{-1.5cm}\mbox{\!\!\!\includegraphics[width=.14\textwidth]{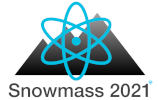}} & &%
\\
 & &  \\   
 &  &  July 29, 2022 \\ 
 & & \\
\end{tabular*}

\vspace*{1.0cm}

{\normalfont\bfseries\boldmath\huge
\begin{center}
  \papertitle 
\end{center}
}

\vspace*{1.0cm}


\begin{center}
\small
{\bf Conveners}: Richard~F.~Lebed$^{1}$, 
Tomasz~Skwarnicki$^{2}$\\
[12pt]
\bigskip
{\it
\footnotesize
$ ^{1}$Department of Physics, Arizona State University, Tempe, AZ 85287-1504, USA\\
$ ^{2}$Department of Physics, Syracuse University, Syracuse, NY 13244, USA
}

\bigskip
\begin{center}
\small
{\bf Contributing Authors}:
Liupan~An$^{3}$,
Sean~Dobbs$^{4}$,
Bryan~Fulsom$^{5}$,
Feng-Kun Guo$^{6,7}$,
Marek~Karliner$^{8}$,
Ryan~E.~Mitchell$^{9}$,
Alessandro Pilloni$^{10,11}$,
Alexis~Pompili$^{12,13}$,
Sasa~Prelovsek$^{14,15}$,
Elena~Santopinto$^{16}$,
Justin~Stevens$^{17}$,
Adam~Szczepaniak$^{18,19,20}$
\end{center}

\bigskip{\it
\footnotesize\noindent
$^{3}$European Organization for Nuclear Research (CERN), Geneva, Switzerland\\
$^{4}$Florida State University, Tallahassee, Florida 32306, USA\\
$^{5}$Pacific Northwest National Laboratory, Richland, Washington 99352, USA\\
$^{6}$CAS Key Laboratory of Theoretical Physics, Institute of Theoretical Physics, Beijing 100190, China\\
$^{7}$School of Physical Sciences, University of Chinese Academy of Sciences, Beijing 100049, China\\
$^{8}$School of Physics and Astronomy, Tel Aviv University, Tel Aviv 69978, Israel\\
$^{9}$Indiana University, Bloomington, Indiana 47405, USA\\
$^{10}$Universit\`{a} di Messina, I-98122 Messina, Italy\\
$^{11}$INFN Sezione di Catania, I-95123 Catania, Italy\\
$^{12}$Dipartimento Interateneo di Fisica, Universit\`{a} degli Studi di Bari, I-70126 Bari, Italy\\
$^{13}$INFN Sezione di Bari, I-70126 Bari, Italy\\
$^{14}$Faculty of Mathematics and Physics, University of Ljubljana, Ljubljana, Slovenia\\
$^{15}$Jozef Stefan Institute, 1000 Ljubljana, Slovenia\\
$^{16}$INFN Sezione di Genova, I-16146 Genova, Italy\\
$^{17}$College of William \& Mary, Williamsburg, Virginia 23185, USA\\
$^{18}$Center for Exploration of Energy and Matter, Indiana University, Bloomington, IN 47403, USA\\
$^{19}$Department of Physics, Indiana University, Bloomington, IN 47405, USA\\
$^{20}$Theory Center, Thomas Jefferson National Accelerator Facility, Newport News, VA 23606, USA\\
}
\end{center}

\vspace{\fill}

\begin{abstract}
  \noindent
  Hadron spectroscopy, the driving force of high-energy physics in its early decades, has experienced a renaissance in interest over the past 20 years due to the discovery of scores of new, potentially ``exotic'' states (tetraquarks, pentaquarks, hybrid mesons, glueballs), as well as the observation of many new ``conventional’’ hadrons. The new discoveries expose our lack of understanding about hadronic states, beyond just a few low excitations of the simplest quark configurations.  
Even so, no single theoretical interpretation (such as hadron molecules, threshold effects, diquark compounds, or others) as yet successfully accommodates all of the new multiquark particles, while a great deal of work remains to extract signals of hybrids and glueballs from reaction-amplitude data.  This document summarizes the current state of the field from both experimental and theoretical perspectives.  On the experimental side, this report summarizes the current, planned, and proposed activities of LHCb and other LHC experiments, Belle II, BESIII, and GlueX, as well as the approved Electron-Ion Collider, the proposed Super Tau-Charm factory, and other future experiments.  The theoretical portion provides a brief overview of multiple phenomenological approaches studied to date, progress in rigorous studies of reaction amplitudes, and advances in lattice-QCD simulations.
\end{abstract}

\vspace*{2.0cm}


\vspace{\fill}

{\footnotesize 
\centerline{\copyright~\papercopyright. \href{\paperlicenceurl}{\paperlicence}.}}
\vspace*{2mm}

\end{titlepage}


\newpage
\setcounter{page}{3}
\mbox{~}


\renewcommand{\thefootnote}{\arabic{footnote}}
\setcounter{footnote}{0}

\tableofcontents
\cleardoublepage


\pagestyle{plain} 
\setcounter{page}{1}
\pagenumbering{arabic}



\section*{Executive Summary}

\begin{itemize}
\itemsep-3pt
    \item Numerous new hadronic states have been discovered in the past 20 years,
      many exhibiting ``exotic'' features that are not compatible with a conventional meson or baryon interpretation.
          No single theoretical model accommodates all the new states. Several different binding mechanisms are likely at play; even mixtures are possible.
      At present, hadron spectroscopy is the least-understood sector of the Standard Model, and thus impacts
      our ability to estimate hadronic effects in BSM searches, to correctly model neutron stars, and to understand the
      spectrum of other strongly coupled theories. 
    \item In the next two decades, the LHC will be the most important facility for hadron spectroscopy, thanks to its high collision energy,
      large strong-production cross sections, and high luminosity. 
      The LHCb upgrades will have the greatest experimental reach, by exploring a wide variety of heavy tetraquark and pentaquark configurations,
      thereby probing hadron structure in different kinematic and dynamical regimes.
      In addition to continued support for LHCb data-taking and analysis,
      investments are needed for detector R\&D, and later for construction of Upgrade-II subsystems. CMS and ATLAS present opportunities for studying final
      states not requiring full hadron identification. 
    \item Belle II complements LHCb via unique access to bottomonium-like vector states and their transitions to other exotics, as well as its
      ability to reconstruct decay modes with multiple neutrals, and to exploit unique production channels.
      Support for detector upgrades is needed to keep up with increasing luminosity, especially after the major SuperKEKB upgrade in a few years.
    \item The BESIII program is the one best poised to explore the anomalous charmonium-like vector states and exotic states seen in their decay, as well as to study
          glueballs in $J/\psi$ decays.
          The proposed Super Tau-Charm factory would increase luminosity by two orders of magnitude and lead to precision studies in these sectors. 
          Participation in data taking and analysis by U.S.\ physicists is strongly recommended.
    \item GlueX, the proposed JLab24 upgrade, and the EIC experiment offer a U.S.\ option for hadron spectroscopy.
          While their electro- and photo-production cross sections and effective collision energies are limiting factors,
          these facilities may succeed in production of numerous heavy-quark exotic states, providing additional insight into their nature.
    \item Theoretical analysis of the new states has been slowed in the U.S. by limited support,
          in contrast to a much broader backing for such research in Europe and Asia. 
\item
    Modern hadron-spectroscopy analysis will require collaborations featuring frequent interactions between experimentalists and theorists,
      and also the collective work of multiple theory researchers.  Lattice-QCD simulations form an essential ingredient of both of these efforts, and their progress requires
      substantial computing resources.
      We therefore call for funding to support consortia of both of these types of group efforts.
      
  \item    Collaboration across nuclear and high-energy communities is also essential for the exchange of expertise and for the development of unified approaches for
      light- and heavy-quark hadrons, and requires a flexible approach from the relevant funding offices.
\end{itemize}

\section{Introduction}
\label{sec:introduction}

As the strongest of all the fundamental interactions, the strong nuclear force has profound consequences for the makeup and behavior of the Cosmos.  Although the Higgs mechanism is touted as the origin of mass for all the Standard Model particles like quarks, the bulk of the mass of protons and neutrons (which, in turn, accounts for almost the entire mass of the visible universe) derives from the binding energy of innumerable {\it gluons\/} (strong-force carriers) that confine together the much lighter quarks.  Exactly how they are assembled remains obscure. Knowing which {\it hadrons\/} (any particles bound by gluons) exist, the pattern of their masses, and what reactions they can undergo---called {\it hadron spectroscopy\/}---is therefore directly related to the most fundamental issues of physics. In spite of nearly a century of experimental and theoretical efforts, our understanding of hadrons remains fragmentary, as revealed by a number of unexpected discoveries of ``exotic'' hadrons made in just the last two decades. This document summarizes the most important questions still to be answered and future prospects to address them. 

Hadron spectroscopy occupies a pivotal role in the history of particle physics.  The explosion of hadron discoveries starting with $\pi^\pm$ in 1947 led directly to the understanding of quarks $q$ as fundamental matter particles in 1964\@.  The proposal of color charge and of gluons as strong-force mediators immediately followed, ultimately pointing to the theory of quantum chromodynamics (QCD) in 1973\@.  Experimental  advances also led to the discovery of the heavier charm $c$ (1974) and bottom $b$ (1977) quarks $Q$ as fundamental denizens of the Standard Model, the spectroscopy of their narrow $Q\bar Q$ bound states ({\it quarkonium}) providing a spectacular vindication of the hypothesis of the quark pair interacting through a simple potential-energy function.

One could be forgiven for thinking that, apart from subsequent experiments simply filling out additional quark-model multiplets of mesons ($q\bar q$) and baryons ($qqq$), the study of hadron spectroscopy no longer lies at the forefront of new discoveries in particle physics.  But in fact, it had been recognized even in the very first quark-model papers~\cite{Gell-Mann:1964ewy,Zweig:1964ruk,Zweig:1964jf} that multiquark ``exotic'' color-singlet hadrons of valence-quark content $q\bar q q\bar q$ (tetraquarks), $qqqq\bar q$ (pentaquarks), {\it etc.}, are allowable.
Perturbative QCD suggests that diquarks ($qq$) in their attractive color-antitriplet configuration could play a role equivalent to a single antiquark $\bar q$, providing a dynamical prescription for building compact multiquark states.  If mesons have sufficiently large binding forces to other mesons or baryons via nuclear-type interactions through the exchange of light-quark pairs, then extended loosely bound ``molecular" states could be realized in multiquark configurations.  In addition, QCD introduced color-octet gluons ($g$) that could serve as valence hadron components, thus giving rise to hybrid ($q\bar q g$, $qqqg$) and glueball ($gg$) states as additional candidate exotics.  Even so, almost a half century elapsed after the development of the quark model in which no unambiguous example of an exotic hadron was observed.

The situation began to change rapidly with the 2003 discovery by the Belle Collaboration~\cite{Choi:2003ue} of a new state $X(3872)$ lying within the mass range of charmonium, but exhibiting properties rather unlike those expected for a pure $c\bar c$ state.  In the two decades since then, $X(3872)$ has been joined by over 50 additional exotic heavy-hadron candidates observed at high statistical significance, most of them being charmonium-like mesons believed to be tetraquarks or hybrids, but also including pentaquarks, bottomonium-like tetraquarks, and open-charm and fully charmed tetraquarks.  The exotic multiquark content of many of these states is obvious from their decay channels, {\it e.g.}, to a quarkonium state and a light charged meson or baryon, with neither of the final hadrons being easily ascribed to low-energy gluon fragmentation.

Studies of heavy-hadron exotics are complemented by studies of light hadrons. Since light $q\bar q$ pairs are easily created from interaction energy, exotic states (including glueballs) mix with ordinary hadrons. Thus, a hunt for light mesons with quantum numbers that cannot be produced in $q\bar{q}$ systems ({\it e.g.}, $J^{PC}=1^{-+}$) plays a special role.  The evidence for other light exotic states usually appears as supernumerary states (beyond the number expected in the conventional spectroscopy), or as an unusual pattern of decay branching fractions hinting at an exotic admixture.

Unlike the situation after the discovery of charmonium, however, no single theory picture has emerged that can accommodate all of the exotic-hadron candidates.  While QCD is believed to be the exact theory of strong interactions, its applications to spectroscopy are limited. Lattice QCD has had great success in describing nearly stable $q\bar q$ and $qqq$ configurations from first-principles QCD\@. However, exact modeling of multiquark states on the lattice is difficult due to their often unstable nature, extended sizes, and the large number of operators needed to capture their full dynamics. Consequently, the most sophisticated theories of exotic hadrons are QCD-motivated models that attempt to identify the most important degrees of freedom. Several competing paradigms ({\it e.g.}, hadronic molecules, di-hadron threshold rescattering effects, direct color-coupling schemes like diquark bound states, the hadrocharmonium model, and others) fail to capture the richness of even just the currently known candidates.  Quite possibly, the eventual universal model of exotic hadrons to be developed over the next decade will turn out to require quantum-mechanical mixtures of more than one of these pictures.  Obtaining this unified scheme will undoubtedly require more sophisticated phenomenological models with input from improved lattice-QCD simulations for multiquark systems, as well as from elaborate, multi-investigator modern approaches to the study of resonant lineshapes in the presence of several decay thresholds.

With multiple major experiments currently running, plus several others in advanced stages of construction or planning, one anticipates scores of additional exotics to be discovered over the next decade. These states include some in novel quark configurations, as well as the refutation of other states currently seen at low statistical significance or predicted by certain models. This document summarizes experimental and theoretical challenges and opportunities in hadron spectroscopy over the next decade, as identified by a diverse group of experimentalists spanning several facilities operating or planned worldwide, as well as by a substantial theoretical community dedicated to phenomenological modeling, amplitude-analysis techniques, and lattice simulations. We provide recommendations for the U.S.-based community, together with its international partners, to maximize its impact on this least-understood part of the Standard Model. Its significance also stretches far beyond hadron spectroscopy, due to the ubiquitous presence of confining strong interactions in beyond-Standard Model searches.

\section{Experimental Landscape and Prospects}

\subsection{The Large Hadron Collider}
\label{sec:lhc-intro}

Hadrons with heavy quarks ($c$ and $b$) play an important role in hadron spectroscopy. Their masses are distinct and large relative to the QCD confinement scale, resulting in easily identifiable hadron multiplets, which often contain at least a few narrow states.  Such narrow states are rather easier to discover and study experimentally, and they provide incisive clues about heavy-quark binding mechanisms. Comparative studies between similar families of hadrons in which a $c$ quark is replaced by a $b$ quark provide additional insight into their internal dynamics. Furthermore, weak $b \! \to \! c$ transitions inside strongly stable hadrons provide an excellent experimental laboratory for hadrons containing a $c$ quark. Since interesting exotic-hadron configurations are often produced with small rates, experiments with high production rates for $b$ and $c$ quarks are required for their studies.
The Large Hadron Collider (LHC) at CERN offers the highest production rates of heavy quarks that will be available to any facility for the next two decades, thanks to large strong-production cross sections, as well as the high instantaneous luminosity of this hadron collider.  The multi-TeV proton-proton collision energy is well above threshold to produce multiple heavy-quark pairs per each beam-beam interaction. However, the collisions also induce potentially enormous backgrounds that must be suppressed not only at the trigger level to keep up with a 40~MHz collision frequency, but also via subsequent offline data analysis, in order to provide meaningful measurements. Therefore, the capabilities of the LHC experiments for hadron spectroscopy are strongly dependent upon their detectors and upon the physical criteria underlying choices for their triggering pathways.  
 
\subsubsection{The LHCb Experiment}
\label{sec:lhcb}

As the first-ever hadron-collider program optimized for heavy-flavor physics, the LHCb experiment~\cite{LHCbCollaboration:2806113} has a number of unique capabilities that enable a broad heavy-hadron spectroscopy program. The LHCb detector covers forward angles, at which substantial heavy-quark cross sections can be captured in a relatively limited solid angle. Its large trigger bandwidth to storage is mostly devoted to heavy-quark physics, unlike for the largest LHC experiments, which optimize to high transverse-momentum ($p_T$) physics. 
With the beam-collision point located at one end of the underground cavern, the full length of the experimental hall can be utilized to deploy additional particle detector technologies. Unlike the high-$p_T$ experiments, LHCb is equipped with two RICH detectors that distinguish between final-state charged hadrons. The ability to distinguish charged kaons originating from the weak decays of heavy quarks, $(b \! \to) c \! \to \! s$, from charged pions (which are the most common product of gluon fragmentation) is important for background suppression. The ability to distinguish protons from kaons and pions is also important for background suppression in final states originating from the decays of heavy baryons.
The Lorentz boost of particle momenta in the forward direction allows LHCb to identify and trigger on muons with lower $p_T$ thresholds than is possible in the central detectors. This feature increases detection efficiency for the final states containing dimuon pairs from the decays of the narrow vector-quarkonia states [$J/\psi$, $\psi(2S)$, $\Upsilon(nS)$], which are important trigger paths for hadron collider experiments. The Lorentz boost also minimizes multiple scattering of charged particles passing through the vertex detector, which helps with the precise determination of secondary vertices produced by weak decays of heavy hadrons. Their separation from the primary beam-beam interaction point serves as the trigger path for final states without muons. 

A combination of the unique properties of the LHCb detector and of the LHC itself resulted in a flood of major discoveries in hadron spectroscopy from the Run 1 and 2 data samples.
The narrow pentaquark states, $P_c(4312)^+$, $P_c(4440)^+$, and $P_c(4457)^+$ decaying to $J/\psi \, p$~\cite{LHCb:2019kea} changed the outlook for multiquark states with baryon number, reversing pessimism about the existence of such states set off in particle-physics community after previous claims of pentaquarks [{\it e.g.}, $\Theta^+(1535) \! \to \! n K^+$] did not survive experimental scrutiny. The $P_c^+$ discovery channel $\Lambda_b\to J/\psi p K^-$ is not accessible at the other facilities expected to operate in the near future. 
While their masses and widths hint at the importance of $\Sigma_c^+ \bar D^{(*)0}$ pairs in their internal dynamics, much remains to be done to clarify the situation. The determination of their quantum numbers, the detection of other decay modes, the search for isospin partners and other production mechanisms, as well as the search for other possible pentaquarks created by the same (loosely bound molecular states?) or other (tightly bound diquark states?) dynamics, require much bigger data samples. In fact, evidence for other pentaquark states, $P_{cs}^0\to J/\psi\Lambda$~\cite{LHCb:2020jpq,LHCb-PAPER-2022-031} and $P_c(4437)^+\to J/\psi \, p$~\cite{LHCb:2021chn}, has been obtained, some with marginal statistical significance and thus needs to be verified. The LHCb Upgrade, now at its commissioning phase, should yield data samples a factor of 10 bigger over the next decade. The upgrade to fully software-trigger with a 40-MHz readout of all detectors will allow LHCb to run at a higher instantaneous luminosity, and yet, still at an order of magnitude smaller than what LHC can already deliver. The LHCb Upgrade II planned for the following decade will take advantage of the full LHC luminosity via use of increased detector granularity, improved radiation hardness, and timing information to cope with the increased number of $pp$ interactions per bunch crossing.

LHCb is most effective in the reconstruction of decay modes containing only charged particles. While $B$-factory experiments like Belle~II (Sec.~\ref{sec:belleii})
have better absolute charged-particle tracking efficiencies, the strong heavy-quark production cross section at LHC more than compensates for that factor.
As a result, LHCb has already accumulated samples orders of magnitude larger than previously available at $B$ factories, which has led to better insight into previously discovered exotic hadrons. Studies of $X(3872)$ in $\pi^+\pi^-\ J/\psi$ ($J/\psi\to\mu^+\mu^-$) are a good example~\cite{LHCb:2013kgk,LHCb:2015jfc,LHCb:2020xds,LHCb:2020fvo,LHCb:2022bly}.
For states produced in $B$ decays [such as in $B^+ \! \to \! X(3872) K^+$], backgrounds are even smaller than at $B$ factories. This reduction is thanks to the large time-dilated increase in $B$-meson decay times in the forward direction at LHCb, which eliminates confusion between the decay products of the two $b$ hadrons, and the remnants of the proton-proton collision. Excellent resolution of primary interaction points and secondary decay vertices protects signal purity against pile-up effects, even at upgraded LHCb luminosities. In fact, the signal statistics in such modes have already reached levels achievable only with the full statistics of upgraded SuperKEKB. LHCb will maintain a large sensitivity advantage through the upgrade programs.\footnote{For example, LHCb detected 6,800 events for $B^+ \! \to \! X(3872)K^+$, $X(3872) \! \to \! \pi^+\pi^- J/\psi$ decays from 9~fb$^{-1}$ \cite{LHCb:2022bly}, while Belle detected 173 events in $0.71$ ab$^{-1}$ (using both $K^+$ and $K_s^0$) \cite{Belle:2011wdj}. Belle~II is projected to detect $40\times173=$ 6,900 events by 2031.
The upgraded LHCb will have 46,100 such events by about the same time, with 276,600 events by the end of Upgrade II.}

The LHCb detector is equipped with an EM calorimeter, and is capable of detecting neutral particles ($\gamma$, $\pi^0\to\gamma\gamma$, $\eta\to\gamma\gamma$, $\omega\to\pi^+\pi^-\pi^0$, {\it etc.}). 
The sensitivity of the present LHCb samples in many modes with charged and neutral particles exceeds those previously attained at $B$ factories by a large factor and, depending upon the mode, may already be at the level achievable at SuperKEKB.\footnote{For example, LHCb detected 
$n=591\pm48$ $B^+\to X(3872)K^+$, $X(3872)\to\gamma J/\psi$ decays from 3 fb$^{-1}$ at 7--8 TeV \cite{LHCb:2014jvf}, while Belle detected 
$n=35.7\pm7.9$ from $0.71$ ab$^{-1}$ \cite{Belle:2011wdj}. Since the background levels are worse at LHCb, we convert these to an equivalent number of background-free events [$n_{eq}=(n/\Delta n)^2$]: $144$ and $20.4$, respectively. Extrapolating to the full 9 fb$^{-1}$ data sample, LHCb has already logged $n_{eq}=576$, while Belle~II is expected to achieve $n_{eq}=40\times20.4=816$ by 2031. LHCb with upgraded an EM-calorimeter can achieve up to $n_{eq}=$ 22,050.}
Since detection efficiency and backgrounds in channels with neutrals are sensitive to pile-up, the EM calorimeter performance will be worse for the upgraded LHCb detector. 
Hadron spectroscopy in LHCb Upgrade II would greatly benefit if the EM calorimeter made use of improved technologies that incorporate precision timing information to suppress pile-up effects.

The LHCb upgrades will advance many other fronts opened by the LHCb discoveries: fully heavy tetraquarks with hidden flavor (the promptly produced structure $X(6900)$ was discovered in the $J/\psi J/\psi$ mass distribution~\cite{LHCb:2020bwg}), doubly flavored tetraquarks (the narrow doubly charmed tetraquark state $T_{cc}^+$ decaying to $D^0D^0\pi^+$ was discovered in prompt production~\cite{LHCb:2021auc}), tetraquark states with open charm and strangeness (two states were discovered in $D^-K^+$ decays, produced via $B^+\to D^+D^-K^+$~\cite{LHCb:2020bls,LHCb:2020pxc},
and others in $D_s^+\pi^+$ ($D_s^+\pi^-$), produced in $B^+\to D^-D_s^+\pi^+$ ($B^0\to\bar{D}^0D_s^+\pi^-$)~\cite{LHCb-PAPER-2022-026}). There is also a variety of tetraquark candidates with hidden-charm and light quarks observed by Belle, BESIII, {\it etc.} and by LHCb in decays to $J/\psi \, \pi^+$, $J/\psi K^+$, $J/\psi \, \phi$, $J/\psi \, \pi^+\pi^-$, $J/\psi \, \eta$, $J/\psi \, \omega$,
$J/\psi \, \phi$, and to charmed meson-antimeson pairs, 
with poorly understood internal dynamics, that needs to be clarified. LHCb has also significantly advanced the spectroscopy of conventional heavy mesons and baryons ({\it e.g.}, discoveries of the narrow families of $\Omega_c$ and $\Omega_b$ states~\cite{LHCb:2017uwr,LHCb:2020tqd}, the doubly heavy baryon $\Xi_{cc}$~\cite{LHCb:2017iph}, and many new baryons with beauty).
Reviewing the details of each of these cases is hardly necessary in this summary document. Given how much has been learned in hadron spectroscopy from the first 10 years of the LHCb program, it is not a speculation to expect many surprise discoveries from the next two decades of the LHCb program. 

Opportunities to study hadron spectroscopy at LHCb add to an already broad scope of research in the LHCb program, with its unique sensitivity reach to BSM physics in rare decays of heavy quarks and in CP-violation studies~\cite{LHCbCollaboration:2806113}. While U.S.\ participation in LHCb has grown somewhat over the years, it is still small relative to that in the other LHC experiments. The latest addition of groups from the nuclear community resulted in unexpected gains in hadron spectroscopy, with the discovery that the $X(3872)$ prompt cross-section dependence on the collision particle multiplicity is different than for $\psi(2S)$~\cite{LHCb:2020sey}. Future production studies in heavy-ion collisions will similarly shed light into the internal structure of this and other exotic hadrons.    
An even larger U.S.\ participation in the LHCb program would result in broader benefits to the American particle-physics community. The Upgrade II program offers detector R\&D and construction opportunities~\cite{LHCbCollaboration:2776420}. As discussed above, an improved EM calorimeter would significantly broaden the scope of final states used to study exotic hadrons, via better detection of neutral particles, in addition to improving electron identification for a more sensitive test of $e/\mu$ universality in loop decays of the $b$ quark.
New tracking stations mounted on the magnet side walls will greatly improve $D^{*+}\to\pi^+D^0$ reconstruction efficiency by improving soft-pion detection, which would benefit many hadron-spectroscopy studies.

\subsubsection{The CMS and ATLAS Experiments}
\label{sec:cms}
The CMS and ATLAS detectors  
cover the central-rapidity region in $pp$ collisions at the LHC, where the heavy-quark cross section is numerically comparable to the one in the forward region covered by the LHCb detector. While the instantaneous luminosity is intentionally dialed down for the LHCb detector, the central detectors have been running at the full LHC luminosity, thus receiving the largest flux of heavy-quark decay products. However, with their main trigger paths devoted to high-$p_T$ physics, only a small fraction of heavy-quark events can be saved for further analysis. The lack of charged-hadron identification devices makes the sensitivity of these detectors to heavy-flavor physics strongly dependent upon particular final states. 

For example, the CMS experiment \cite{Chistov:2022rht} 
has a better sensitivity than LHCb to prompt production of $\Upsilon\Upsilon$ meson pairs detected in the $\mu^+\mu^-\mu^+\mu^-$ final state, and in the related search for $b\bar{b}b\bar{b}$ tetraquarks decaying to $\Upsilon\mu^+\mu^-$ \cite{CMS:2016liw,CMS:2020qwa,LHCb:2018uwm}.
This feature is thanks to its large-acceptance muon detector, and the sufficiently large $\Upsilon$ mass to produce muons above the CMS $p_T$ trigger thresholds.
Investigating this sector with larger data samples in the future will be important in view of the LHCb discovery of $J/\psi J/\psi$ mass structures \cite{LHCb:2020bwg}. 
Verifying this claim, and looking for $b\bar{b}c\bar{c}$ states in the $\mu^+\mu^-\mu^+\mu^-$ final state, is also an important task for the central detectors. 
In fact, the recent preliminary results from CMS~\cite{CMS-PAS-BPH-21-003} 
and ATLAS~\cite{ATLAS-CONF-2022-040}
not only confirm the $X(6900)$ structure observed by LHCb, but also point to at least two additional mass peaks~\cite{CMS-PAS-BPH-21-003} and a structure in the $J/\psi \, \psi(2S)$ mass~\cite{ATLAS-CONF-2022-040}.

No hadron-identification devices are needed for tagging strangeness via the reconstruction of $K_s^0\to\pi^+\pi^-$ or $\Lambda\to p K^-$, which create narrow mass peaks for oppositely charged track pairs, as well as produce well-isolated secondary vertices.
The forward-momentum boost makes many of them decay beyond the extent of the vertex detector in LHCb, diluting the related sensitivity. This is less of a factor for their detection, since $\Lambda$ and $K^0_s$ can be identified in the higher-level trigger in the central region.  Therefore, CMS is likely to play a role for final states with $J/\psi\to\mu^+\mu^-$ and involving $\Lambda$ or $K_s^0$. 
For example, CMS inspected the $B^+\to J/\psi p \bar{\Lambda}$ final state for possible pentaquark signals \cite{CMS:2019kbn}. 
The $\Lambda_b\to J/\psi p K^-$ channel, in which LHCb discovered the $P_c^+\to J/\psi p$ states \cite{LHCb:2019kea}, is difficult for the central detectors because of the large backgrounds without hadron identifications, as illustrated by the ATLAS analysis \cite{ATLAS:2019keh}.

Even without charged-kaon identification, significant background suppression can be achieved in channels with $J/\psi\to\mu^+\mu^-$ and $\phi\to K^+K^-$, since the $\phi$ mass peak is narrow and near threshold. In fact, the CMS experiment confirmed the $X(4140)\to J/\psi\phi$ state in $B^+\to J/\psi\phi K^+$ decays \cite{CMS:2013jru}, first claimed by the CDF detector at the Tevatron \cite{CDF:2009jgo}, before the LHCb experiment investigated this channel more thoroughly \cite{LHCb:2016axx,LHCb:2021uow,LHCb:2016nsl}. The first measurement of the $B_s\to X(3872)\phi$ branching ratio by CMS 
\cite{CMS:2020eiw} is another example [$X(3872)\to\pi^+\pi^- J/\psi$].
The detection of $\Lambda_b \to J/\psi\Lambda\phi$ by CMS combines benefits of the reconstruction of $\Lambda$ and of $\phi$ \cite{CMS:2019mny}. 

There are many more pions produced promptly in the forward direction than perpendicularly to the beams, which offers a competitive edge to the central detectors in studying soft-pion transitions of promptly produced excited $B_c$ states \cite{ATLAS:2014lga,CMS:2019uhm,LHCb:2019bem,CMS:2020rcj} or $b$ baryons, to the respective ground-state levels of these systems, which are detected in channels with $J/\psi \to\mu^+\mu^-$. 

The recent detection of the $X(3872)\to\pi^+\pi^- J/\psi$ signal in Pb-Pb collisions by CMS hints at enhanced production of this state in heavy-ion collisions relative both to $\psi(2S)$ and from $pp$ collisions \cite{CMS:2021znk}. If confirmed, this result would have significant implications for interpretations of the $X(3872)$ state, and illustrates a possible new frontier for studies of exotic-hadron candidates.  Analogous studies are also possible at ATLAS~\cite{ATLAS:2022hsp}; more data are needed. 

While the number of final states in which the central detectors at the LHC can have sensitivity better than that of the LHCb is limited, some of them, like tetraquarks decaying to $\Upsilon$, may lead to major new discoveries. The  installation of timing-layer detectors for CMS in the upcoming upgrade for the HL-LHC phase will be particularly helpful in suppressing backgrounds.  In channels in which LHCb is likely to dominate, it is still very important to verify LHCb claims whenever possible, especially for those channels not accessible at Belle~II ({\it e.g.}, exotic hadrons with multiple heavy flavors, or produced in decays of $b$ baryons). Since these priorities are set at the trigger level and by the allocation of computing resources, careful planning for future data taking is necessary and should receive appropriate support.

\subsubsection{The ALICE Experiment}
\label{sec:alice}

For a fraction of each year, the LHC circulates and collides heavy ions instead of protons. The LHCb, CMS, and ATLAS experiments take data during that time, and each has a dedicated heavy-ion data-analysis group. One of the four beam-collision points is occupied by the dedicated heavy-ion experiment, ALICE\@. ALICE has been able to study charmonium production in heavy-ion collisions (see, {\it e.g.},~Ref.~\cite{ALICE:2022sli}). 
While optimized for high-multiplicity event reconstruction, the Runs 1 and 2 ALICE detector was not optimal for the detection of heavy-quark hadrons,\footnote{Their production cross sections are suppressed by the heavy-quark mass.} since its main TPC tracking detector was too slow to operate at high readout frequency. 
In fact, CMS (Sec.~\ref{sec:cms}) was the first experiment to detect $X(3872)$ in heavy-ion collisions. The TPC readout speed has been upgraded for the upcoming Runs 3 and 4 of the LHC in the next decade~\cite{ALICETPC:2020ann}, 
but it will be still a limiting factor.
A new detector, ALICE 3, is being proposed for Run 5 in the following decade. With an all-silicon based tracking system~\cite{Adamova:2019vkf,Colella:2022lmb}
and a large-barrel RICH detector, such an experiment would have a much increased sensitivity to heavy-quark exotics~\cite{ALICE3}.

\subsection{The Belle~II Experiment}
\label{sec:belleii}

The $e^+e^-$ $B$ factories, the Belle experiment in particular, played a key role in pushing meson spectroscopy beyond the quark-antiquark paradigm. The discovery of the $X(3872)$ state \cite{Choi:2003ue}, of the $Y(4260)$ structure \cite{BaBar:2005hhc}, and of the $Z_c(4430)^+$ state \cite{Belle:2007hrb} marked the beginning of what is often called the $XY \! Z$ era. These events were followed by many other observations of heavy exotic-meson candidates at the $B$-factory experiments (see, {\it e.g.}, Table I in Ref.~\cite{Olsen:2017bmm}).      

Belle~II is the next-generation $B$-factory experiment at the SuperKEKB $e^+e^-$ collider in Japan operating near the $b\bar{b}$ threshold. Thanks to the novel concept of focused ``nanobeams" at the interaction point, SuperKEKB is designed to deliver a much higher instantaneous luminosity than was previously available. Over the next decade, Belle~II is expected to collect 40 times more data than analyzed by the Belle or BaBar experiments. Since the main goal of the experiment is to search for BSM physics in decays of $B$ mesons, most of the data is being taken at the $\Upsilon(4S)$ resonance, which maximizes the $B^+B^-$ and $B^0\bar{B}^0$ production rates. $B_s^0$ mesons can be produced as well, albeit with a lower cross section, by dedicated runs at the $\Upsilon(5S)$ resonance. The $\Lambda_b \bar{\Lambda_b}$ threshold could be also reached after modifications of the storage ring, but this enhancement is not included in the currently approved program. 

The overall goals of the Belle~II and LHCb programs are similar, and center around BSM searches in decays of heavy quarks. As such, they are also excellent experiments to study the spectroscopy of heavy hadrons. While there is an overlap in the programs, there are also complementary domains for each experiment.  The LHCb program benefits from larger heavy-quark production cross sections (strong {\it vs.\ }electromagnetic couplings), holding advantages for all-charged-particle final states and some simple channels including a neutral particle (Sec.~\ref{sec:lhcb}). On the other hand, Belle~II is competitive in channels with multiple neutral particles, and offers a number of unique areas of sensitivity because of different production mechanisms specific to an $e^+ e^-$ collider.
Perhaps the most important one is the ability to operate at a tuneable center-of-mass $e^+e^-$ energy just above the $B\bar{B}$ threshold, in the search for $b\bar{b}q\bar{q}$ tetraquarks or $b\bar{b}g$ hybrids.
In fact, the data taken at $\Upsilon(5S)$ led to the discovery of the narrow $J^P \! = \! 1^+$ $Z_b(10610)^+$, $Z_b(10650)^+$ states near the $B\bar{B}^*$ and $B^*\bar{B}^*$ thresholds, respectively,
decaying to $\Upsilon(nS)\pi^+$ ($n=1,2,3$) and $h_b(mP) \pi^+$ ($m =1,2$)~\cite{Belle:2011aa}, explicitly identifying them as 4-quark effects.  
Together with their isospin partners, these are the only undisputed exotic hadrons containing $b$ quarks. 
The scan of $\Upsilon(nS)\pi^+\pi^-$ cross sections revealed a possible $J^{PC} \! = \! 1^{--}$ $Y_b(10753)$ resonance~\cite{Belle:2019cbt}, which could either be an $\Upsilon(3^3D_1$) $b\bar{b}$ state, a hybrid, or a tetraquark.
Since similar scans above the $c\bar{c}$ threshold produce a rich spectrum of exotic phenomena in various charmonium-plus-light-hadrons channels, more precise scans of the $b\bar{b}$ sector would likely reveal more states, or at least provide additional input to the interpretation of the states with hidden charm. None of the charmonium-like or bottomonium-like states of this type have been established in prompt production at hadron colliders. Thus, only the Belle~II program has access to such data, assuming dedicated runs to search for them. 

There are other production mechanisms unique to Belle~II that do not require dedicated runs. Among them, $\gamma\gamma$ collisions reach the experiment's sensitivity threshold for the detection of charmonium-like states. Discoveries of new states are certainly possible, but negative search results for states observed elsewhere are also meaningful, by providing constraints upon their viable interpretations. The other unique mechanism at Belle~II is double-charmonium production in continuum $e^+e^-$ annihilation. Studying the recoil mass spectrum against the reconstructed $J/\psi$ meson produced a new, possibly exotic structure $X(3940)$~\cite{Belle:2005lik}. Combining the $J/\psi$ with a reconstructed $D$ or $D^*$ meson allows to disentangle contributions from states decaying to $D^{(*)}\bar{D}^{(*)}$ channels \cite{Belle:2007woe}. Increasing statistics in these data sets, as well as employing this method with the recoil against the $\psi(2S)$, $\eta_c$, or $\chi_{cJ}$ mesons, will be very interesting.

The initial-state radiation (ISR) process lowers the effective $e^+e^-$ collision energy, allowing access to the $c\bar{c}$ threshold region.
Since no dedicated runs are required, and all masses are probed simultaneously, this production mechanism provides for a useful survey of a wide range of collision energies, and has led in the past to a number of important discoveries \cite{BaBar:2005hhc,Belle:2007umv,Belle:2013yex}. Scanning regions of interest above the $D\bar{D}$ threshold at the BEPCII tau-charm $e^+e^-$ factory provided for more sensitive probes of this sector (see Sec.~\ref{sec:BESIII}). Both experimental approaches produce corroborating results~\cite{Belle:2013yex,BESIII:2013ris}. With a 50~ab$^{-1}$ sample, Belle~II may reach better sensitivity in some channels until a super tau-charm factory is built (Sec.~\ref{sec:futuretaucharm}). 
However, since energy scans require dedicated runs at tau-charm factories, the $B$-factory results provide a useful guide for them.

Belle~II also has unique access to conventional-bottomonium spectroscopy. Many previously undiscovered $b\bar{b}$ states were observed via hadronic transitions between them, when taking data above the $B\bar{B}$ threshold  \cite{Belle:2015tbu}. Belle~II has the potential to discover $\Upsilon(2D)$ and $\Upsilon(1F)$ states \cite{BelleIIwhitepaper} as well.

Belle~II began its physics run in 2019, and so far has accumulated 0.4~ab$^{-1}$ of integrated luminosity. SuperKEKB set a new instantaneous luminosity record of $4.1 \! \times \! 10^{34}$~cm$^{-2} \, $s$^{-1}$,
exceeding the old KEKB record of $2.1 \! \times \! 10^{34}$~cm$^{-2} \, $s$^{-1}$. 
Further improvements of up to $2 \! \times \! 10^{35}$~cm$^{-2} \, $s$^{-1}$ are expected using the existing accelerator complex. However, to reach the integrated luminosity goal of 50 ab$^{-1}$ by 2031, 
an upgrade of the interaction region to deliver the instantaneous luminosity of $6.5\times10^{35}$~cm$^{-2} \,$s$^{-1}$ will likely be needed around 2026. Discussions about a possible extended program into the following decade has started, with a luminosity upgrade up to $2\times10^{36}$~cm$^{-2} \,$s$^{-1}$. 
The Belle~II detector is performing well. The second layer of the pixel vertex detector will be installed during the long shutdown that starts in summer 2022. Photodetector and readout upgrades of hadron-ID devices are planned for the next long shutdown in 2026\@. A new vertex detector may also be required, due to the redesign of the interaction region \cite{Asner:2022axe,Forti:2022mti,Natochii:2022vcs}.

As noted previously, searches for BSM physics in decays of heavy quarks are the main goal of the Belle~II program.  However, as outlined above, Belle~II is also expected to play an important role for hadron spectroscopy. Some of the physics reach is unique, and some is complementary to that of the LHCb and BESIII programs. U.S.\ participation in the Belle~II program is modest, and at a scale similar to that in LHCb. Both provide excellent physics return for the amount of investment made.  Advancing hadron spectroscopy calls for continued U.S.\ involvement in the data processing and analysis of the experimental results, as well as extensions to the beam energy to produce collisions above $\Upsilon (4S)$ to 11~GeV and beyond.

\subsection{The BESIII Experiment}
\label{sec:BESIII}

The BESIII experiment at the BEPCII Collider in China, which currently accesses $e^+ e^-$ collisions at c.m.\ energies 2--5~GeV, is the world's premiere facility for precision $\tau$/charm physics~\cite{BESIII:2022mxl}.  
It operates near the $c\bar{c}$ threshold, and its role in hadron spectroscopy is analogous to the one played by $B$ factories operating near the $b\bar{b}$ threshold.
In its lower energy range, BESIII can operate near the $\phi\phi$ threshold.
Relative to Belle II, studies of hadrons with charmed quarks are easier at BESIII due to the lower multiplicity of events, the threshold kinematics, and the larger cross sections. 
However, BEPCII instantaneous luminosity is lower than that of SuperKEKB, and has reached $1\times10^{33}$~cm$^{-2} \, $s$^{-1}$.
From its 2009 inception to date, BESIII has collected 35~fb$^{-1}$ of data.
  
Perhaps the most dramatic hadron results from BESIII arise from scans over collision energies above the $D\bar{D}$ threshold, and observations of states not accommodated within the simple $c\bar{c}$ model. 
Since the beam energy can be tuned precisely, and large signal statistics in various channels can be accumulated quickly thanks to the larger cross sections and better reconstruction efficiency, this method produces more accurate determinations of resonant shapes in various decay channels than can be accomplished by producing them via initial-state radiation (ISR) at $B$ factories (Sec.~\ref{sec:belleii}). 
The best illustration of this feature is the resolution of the $Y(4260)$ structure discovered at $B$ factories into two finer structures by BESIII\cite{BESIII:2016bnd}. The $1^{--}$ sector has been investigated  in channels such as $e^+ e^- \! \to \! \pi^+ \pi^- J/\psi$~\cite{BESIII:2016bnd}, $\pi^+ \pi^- h_c$~\cite{BESIII:2016adj}, $\omega \chi_{c0}$~\cite{BESIII:2014rja}, $\pi^+ D^- D^{*-}$~\cite{BESIII:2018iea}, and $\pi^+ \pi^- \psi(2S)$~\cite{BESIII:2017tqk,BESIII:2021njb}.   This study reveals complicated lineshapes that may be interpreted in terms of several new states, traditionally called $Y$ (the new PDG convention labels them by $\psi$).  The lightest of these, $Y(4230)$, has been observed in multiple channels, but other $Y$ states observed by BESIII [{\it e.g.}, $Y(4390)$, $Y(4660)$] give different mass and width measurements in different decay channels.  Whether they are compact resonances (tetraquarks or hybrids) or hadron-rescattering 
effects remains unanswered; BESIII plans a detailed scan between 4.0--4.6~GeV at 10~MeV intervals with 0.5~fb$^{-1}$ per point in order to attempt to resolve such issues.

By running above the $D\bar{D}$ threshold, BESIII has also led the way in the discovery of several hidden-charm, isospin-1 exotic candidates, collectively called $Z_c$.  The first, $Z_c(3900)$, was discovered in the process $e^+ e^- \! \to \! \pi^\pm Z_c^\pm$, $Z_c^\pm \! \to \! \pi^\pm J/\psi$~\cite{BESIII:2013ris}, and $Z_c(4020)$ was discovered in $Z_c^\pm \! \to \! \pi^\pm h_c$~\cite{BESIII:2013ouc}.  Both were subsequently seen by BESIII in open-charm channels~\cite{BESIII:2013qmu,BESIII:2013mhi}, and their neutral partners were also observed at BESIII~\cite{BESIII:2014gnk,BESIII:2015cld}.  In addition, a newly observed near-threshold structure in $e^+ e^- \! \to \! K^+ (D_s^- D^{*0} + D_s^{*-} D^0)$ called $Z_{cs}^-(3985)$~\cite{BESIII:2020qkh} is a candidate $c\bar c s\bar u$ exotic.  Whether $Z_c(3900)$, $Z_c(4020)$, and $Z_{cs}(3985)$ are confirmed as true resonances ({\it i.e.}, appearing with the same measured properties in all channels) that fill complete spin and flavor multiplets, or whether they are peculiar hadron-rescattering effects, 
requires further information.  To address these issues, BESIII plans to collect $O$(5~fb$^{-1}$) at several points near each relevant threshold, in order to perform suitable amplitude analyses.

The state $X(3872)$ has been extensively scrutinized at multiple experiments, but BESIII has contributed significantly to its study through the observation of channels such as $e^+ e^- \! \to \! \gamma X(3872)$~\cite{BESIII:2013fnz}, $X(3872) \! \to \! \pi^0 \chi_{c1}(1P)$~\cite{BESIII:2019esk}, and $X(3872) \! \to \! \omega J/\psi$~\cite{BESIII:2019qvy}.  With a larger data set and extended reach from its upcoming upgrades, BESIII will address questions such as whether $X(3872)$ has any $J^{++}$ partners, either produced by radiative processes or through hadronic transitions of the form $e^+ e^- \! \to \! \omega X$ or $e^+ e^- \! \to \! \phi X$.

BESIII has also contributed very significantly to 
studies of conventional $c\bar c$ states, both below and above the $D\bar{D}$ threshold ({\it e.g.}, the state $\psi (1 {}^3 D_2)$ was first observed with $> \! 5\sigma$ significance by BESIII~\cite{BESIII:2015iqd}). These studies and those of the $XY \! Z$ states are complementary, not only because some of these states might be mixtures of the two types, but also because transitions between them will be crucial to understand their underlying structure.  For example, in the case of $X(3872)$ [which potentially mixes with $\chi_{c1}(2P)$], studying the nearby $\chi_{cJ}(2P)$ and $h_c(2P)$ states will provide critical information.  The lighter $\eta_c$ and $h_c$ states, well studied at BESIII, provide additional probes as $XY\! Z$ decay products.  In addition, transitions between exotics can indicate relations between them; in this case, the process $e^+ e^- \! \to \! \gamma X$~\cite{BESIII:2013fnz} provides evidence for $Y(4230) \! \to \! \gamma X(3872)$, and $e^+ e^- \! \to \! \pi^0 \pi^0 J/\psi$~\cite{BESIII:2015cld} likewise points to $Y(4230) \! \to \! \pi^0 Z_c^0(3900)$, suggesting a common structure between these $XY\! Z$ states.

BESIII collected record-breaking data sets at the narrow charmonium states, including $10^{10}$ $J/\psi$ and $2.7 \! \times \! 10^9$ $\psi(2S)$ decays. Decays of these resonances open a window to light-hadron spectroscopy, both conventional and exotic.  In particular, the annihilation decays of $c\bar c$ states provide a gluon-rich environment in which to produce glueball and hybrid states (such as the recently observed candidate $\eta_1(1855)$~\cite{BESIII:2022riz}).  

The BESIII program is anticipated to continue for another 5--10 years, and approved upgrades will increase the maximum energy of BEPCII to 5.6~GeV, followed by a luminosity increase by a factor of 3\@.
These improvements will offer opportunities to produce pentaquark states in a variety of channels, such as $e^+ e^- \! \to \! J/\psi \, p + \! X$, $\chi^{\vphantom\dagger}_{cJ} \, p + \! X$, $J/\psi \, \Lambda + \! X$,
$\bar D^{(*)} p + \! X$, and $D^{(*)} p + \! X$. Such studies could not only offer confirmatory evidence of the pentaquark candidates seen at LHCb (Sec.~\ref{sec:lhcb}), but also permit the exploration of numerous channels not easily studied elsewhere.

Current U.S.\ involvement in BESIII is limited to a small number of physicists. An investment in funding to support such research provides
an avenue for the U.S.\ particle-physics 
community to engage in an important area of physics not otherwise represented in this country.

\subsection{Super Tau-Charm Factories}
\label{sec:futuretaucharm}

Next-generation tau-charm-factory proposals aim at overcoming the main limitations of the present BESIII/BEPCII program: luminosity and maximal collision energy.
Various sites have been proposed in China \cite{Luo:IPAC2018-MOPML013} and in Russia \cite{Barnyakov:2020vob}.
The instantaneous luminosity would be improved by two orders of magnitude, to about $10^{35}$ cm$^{-2}$s$^{-1}$ at 4 GeV. 
The energy reach would be extended from 5.6 to 7~GeV\@. 
The proposals are at the R\&D stage. If approved, the facility would come into operation in the following decade, with a possibility of further upgrades of luminosity in the farther future.   

Energy scans above the $D\bar{D}$ threshold with the increased luminosity at a super tau-charm factory would significantly impact hadron spectroscopy via~\cite{Guo:2022kdi}:
\vspace{-6pt}
\begin{itemize}
\itemsep-3pt
    \item Producing the most precise studies of directly produced charmonium and charmonium-like vector states ({\it i.e.}, $\psi$ and $Y$ states). Their lineshapes as well as decay patterns would be mapped out, including many new decay modes.
    \item At the same time, other charmonium-like states produced via hadronic and electromagnetic transitions can be similarly studied [{\it e.g.}, $Z_c(3900)$ or $X(3872)$], and their connection to $Y$ states can be better established. 
    \item Both of these advances will allow better insight into $c\bar{c}q\bar{q}$ tetraquarks, as well as searches for $c\bar{c}g$ hybrids.
    \item Developing a better understanding of highly excited $c\bar{c}$ states, and their mixing with exotic hadrons of the same quantum numbers.
\end{itemize}
\vspace{-6pt}
Increased instantaneous luminosity will also lead to much larger samples accumulated atop the narrow charmonium resonances [$3\times10^{12}$ $J/\psi$ and $5\times10^{11}$ $\psi(2S)$], allowing for unique insights into light-hadron spectroscopy \cite{Guo:2022kdi}:
\vspace{-6pt}
\begin{itemize}
\itemsep-3pt
     \item Measurements of electromagnetic couplings of glueball candidates [{\it e.g.}, $f_0(1710)$], which are of key importance for their discrimination from $q\bar{q}$ states.
     \item A survey of decay patterns of light hybrid candidates [{\it e.g.}, $\eta_1(1855)$] to clarify their internal structure and to search for hybrid supermultiplets.
     \item Unique data sets to study light baryons, including the possibility to search for $sssq\bar{q}$ pentaquarks.  
\end{itemize}
\vspace{-6pt}
Extending the maximal energy reach will open new opportunities:
\vspace{-6pt}
\begin{itemize}
\itemsep-3pt
    \item Searches for $P_c$ and $P_{cs}$ pentaquarks in several final states, in both hidden- and open-charm cases.
    \item Searches for exotic states decaying to $\Lambda_c^+\bar{\Lambda}_c^-$ (charmed-baryon or -meson molecules, hexaquarks).
    \item Searches for double-charmonium states in $J/\psi \, J/\psi$, $J/\psi \, \eta_c$, and $J/\psi \, \chi_{cJ}$ final states.
\end{itemize}
\vspace{-6pt}
Given how much the BESIII program has contributed to hadron spectroscopy (Sec.~\ref{sec:BESIII}), increasing the luminosity at the tau-charm factory by two orders of magnitude is guaranteed to produce new discoveries, as well as provide more systematic information on the known states, which will lead to a deeper understanding of the dynamics of exotic states with charm quarks.
It should also clarify evidence for light glueballs and hybrid states.
Participation of U.S.\ groups in this undertaking would be well justified by its physics reach.

\subsection{Photoproduction Experiments}
\label{sec:Photo}
The $pp$ and $e^+ e^-$ collider experiments described in the previous subsections have led the way in the discovery of all exotic-hadron candidates in the heavy-quark sector to date.  However, many of these states are seen in only one production mechanism.  Photoproduction (including electroproduction) at $\gamma^{(*)}p$ or $\gamma^{(*)} \! A$ facilities could provide an alternative for studying such hadrons, both in terms of the range of states accessible and in the complementary nature of the kinematics~\cite{Albaladejo:2022dwv}.  Light-hadron spectroscopy, both conventional and exotic, has been under investigation at photoproduction facilities for a while. 
In particular, real and virtual photons present clean probes of the internal structure of hadrons, because they induce (quasi-)two-body processes, the only independent kinematical variables being the exchange momentum and the photon virtuality.  Photoproduction processes evade the complications of 3-body rescattering effects (particularly near thresholds) that plague the analysis of many of the $XY\! Z$ states, and also allow for access to both low- and high-energy production in the same experiment. While none of the heavy-quark exotic hadrons have been established in photoproduction so far, their observation in this production mechanism would provide an important new tool to probe their internal structure. Experimentally, the task is challenging because the cross sections are limited by the electromagnetic nature of the process, and the need for the photon-hadron collision energy to exceed the heavy-quark pair-production threshold. Since $J/\psi$ photoproduction has been already observed~\cite{GlueX:2019mkq}, the approach holds good promise to produce at least compact charmonium-like exotics (compact tetra- and penta-quarks, or hybrids).  The ability to photoproduce molecular states is more difficult, and predictions are more model dependent. Increasing the high-energy photon flux, either via its intensity or the beam energy, are the main routes for improving sensitivity. 

\subsubsection{The GlueX Experiment}

The GlueX Experiment at Jefferson Lab (JLab) began taking data in 2017, and features a linearly polarized photon beam with lab-frame energies up to 12~GeV, as well as a solenoidal spectrometer with near-full acceptance for both charged and neutral particles~\cite{GlueX:2020idb}.  To date, 840~pb$^{-1}$ of data at beam energies $> \! 6$~GeV has been collected (Phase~I).  Phase~II began in 2020 with a higher photon-beam intensity, as well as the addition a DIRC detector to improve $K$/$\pi$ separation, and is expected to conclude in 2025, by which point almost 3 times more data will be collected.  In addition, an upgrade of the inner portion of the forward calorimeter by incorporating lead tungstate crystals in order to achieve better energy and position resolution is expected in 2023.  JLab is currently the principal facility in the U.S.\ performing hadron spectroscopy and structure experiments.

The original focus of GlueX is the search for light-quark hybrid mesons, especially ones with exotic quantum numbers not accessible to $q\bar q$ states (such as the $J^{PC} \! = 1^{-+}$ $\pi_1(1600)$ and its potential isosinglet partner(s) $\eta_1^{(\prime)}$ very recently observed by BESIII~\cite{BESIII:2022riz}).  However, GlueX also presents opportunities for the study of hidden- and open-strange hadrons, including possible strange-quark analogues to $XY\! ZP$ states~\cite{Hamdi:2019dbr} [such as the candidate $\phi(2170)$], as well as for the photoproduction of charmonium (chiefly $J/\psi$~\cite{GlueX:2019mkq}, but also $\psi(2S)$, off the proton), which offers an alternative production mechanism for the $P_c$ pentaquark candidates first observed by LHCb (see Sec.~\ref{sec:lhcb}).  Such searches are particularly interesting because none of the $XY \! ZP$ candidates have yet been unambiguously observed in electro- or photoproduction experiments.  Once such a state is discerned, however, then collecting data at its resonance pole while scanning over different values of photon virtuality (note that GlueX currently uses only real photons) will provide incisive information about the spatial extent of the resonance ({\it e.g.}, a large hadronic molecule {\it vs.\@} a compact multiquark state).  Measuring polarization dependence provides access to additional observables.

The existence of hybrid mesons (for which the gluon field carries a nontrivial part of the valence quantum numbers) have been studied for decades in both model and lattice-QCD calculations.  However, hybrids in the light-quark sector are expected to mix with conventional mesons of the same $J^{PC}$, thereby complicating their observation; and even ones with exotic $J^{PC}$ have proven tricky to extract from the data.  For example, $\pi_1(1600)$ is expected to decay dominantly to $b_1(1235) \pi$, which ultimately leads to a 5$\pi$ final state.  Alternate channels that are cleaner but have lower branching fractions ({\it e.g.}, $\eta \pi$) are also being examined at GlueX~\cite{Arrington:2021alx}.  Disentangling such complicated reaction amplitudes while imposing core field-theory requirements such as unitarity, causality, and crossing symmetry requires significant theoretical input, such as that from the JPAC Collaboration (which works closely with GlueX as well as with other experimental groups), as described in Sec.~\ref{sec:JPAC}.

GlueX, particularly in Phase~II, will produce a large data set of two-pseudoscalar ({\it e.g.}, $K_S K_S$, $\eta \eta^\prime$) and vector-pseudoscalar ({\it e.g.}, $\phi \pi$, $K_1 K$) final states to which the amplitude-analysis framework discussed above can be profitably applied.  A number of these final states can be investigated in order to search for strange analogues of $XY \! ZP_c$, such as through $Z_s \! \to \! \phi \, \pi$ and $P_s \! \to \! \phi p$, and the possible $Y(4230) \! \to \! J/\psi \, \pi^+ \pi^-$ analogue $\phi(2170) \! \to \! \phi \, \pi^+ \pi^-$~\cite{Hamdi:2019dbr}.

At the highest GlueX photon energy (12~GeV), the photoproduction of charmonium states up to $\psi(2S)$ is possible, although in increasingly small numbers for higher excitations.  GlueX Phase~I is expected to produce about 2000 $J/\psi$'s, 50 $\chi_{c1}$'s, and $< \! 10$ $\psi(2S)$'s, and GlueX Phase~II will increase these yields by about a factor of 3\@.  While these samples are small, they will provide the first studies of the energy dependence of their photoproduction cross sections.  As noted above, GlueX has obtained evidence for $J/\psi$ photoproduction~\cite{GlueX:2019mkq}, even after collecting only about 25\% of the expected Phase-I data.  This study also provides a direct search for $P_c$ states decaying to $J/\psi p$ states; their absence in the data analyzed to date already sets a model-independent bound for a branching fraction of less than a few percent.

The GlueX experiment was approved on the basis of its light-hadron spectroscopy program. The original goals remain valid, and call for completion of its data-taking program in the next few years. Its ability to study photoproduction of charmonium states became important in view of the new particle zoo of charmonium-like states above the open-flavor threshold, including pentaquark states, discovered at other facilities.
Even if the sensitivity of this experiment does not reach the sensitivity required to detect exotic states, the effort constitutes an important step towards that goal.

\subsubsection{Proposed CEBAF Upgrade}

A proposed energy upgrade of the CEBAF Accelerator at JLab using Fixed Field Alternating Gradient in the existing recirculation arcs, called JLab24~\cite{Arrington:2021alx}, would increase the accessible photon energy to 20--24~GeV\@.  Photoproduction of all charmonium(-like) states up to 5.5~GeV (and not just baryonic $P_c$ states) would then become possible. In addition, beam and target polarizations would be adjustable, providing access to multiple new observables.

This proposal would greatly increase the chance of detecting exotic charmonium-like states via photoproduction in an experiment optimized towards a such goal. A successful observation would open a new front in studies of such states.  

\subsubsection{The Electron-Ion Collider}

Construction of the Electron-Ion Collider (EIC), to be located at Brookhaven National Laboratory (BNL) and jointly operated by BNL and JLab, is slated to begin in 2024, with data collection to commence around 2030\@.  The EIC is a high-luminosity polarized $ep$ and $eA$ collider, with variable center-of-mass energies from 20--140~GeV, and it represents the highest-energy machine currently approved for development in the U.S\@. 

The EIC will have sufficient energy to produce essentially all known $XY \! ZP$ candidates, including both the charmonium and bottomonium energy ranges~\cite{AbdulKhalek:2021gbh}.  Indeed, its higher energy is ideal for studying diffractively produced $Y$ states, whose cross sections are expected to rise significantly with energy.  With an integrated luminosity of $O(1$-$10~{\rm fb}^{-1})$, the EIC will present unique opportunities to perform detailed studies of numerous rare exclusive processes for the first time.  In addition, the ability to vary photon virtuality in photoproduction processes permits measurements of the spatial extent of the states, which provides crucial information on their substructure.

Using photoproduction to determine the spectroscopy of exotic quarkonia requires high detection efficiency for all the decay products.  Detailed studies have been carried out to examine sample reaction chains, such as $\gamma p \! \to \! Z_c(3900)^+ n$, $Z_c(3900)^+ \! \to J/\psi \, \pi^+$, $J/\psi \! \to \! e^+ e^-$~\cite{Glazier:2020}.  From a theoretical point of view, an in-depth amplitude analysis such as that described in Sec.~\ref{sec:JPAC} must be incorporated.  Furthermore, the asymmetric collider kinematics of the EIC requires a specialized detection system to identify not only the decay products of the created states---which may be quite boosted in the proton-beam direction and require detector acceptance up to high pseudorapidity values ($\approx$~3.5)---but also to tag the scattered $e^\prime$ and $p^\prime$ in order to reconstruct the full exclusive process, and to efficiently reconstruct subprocesses such as $J/\psi \! \to \! e^+ e^-$.  Nevertheless, comprehensive studies such as described above for $Z_c(3900)^+$ predict clean signal-to-background ratios, and provide encouraging indications for the future discovery potential of the EIC\@.

The EIC main goals are precision studies of structure functions of nucleons and nuclei. As a side product, thanks to its large luminosity and high beam energy, the EIC may be able to contribute to the spectroscopy of charmonium-like states. Exact predictions of how much can be accomplished await first observations of such exotic states in photoproduction. If successful, their studies at the EIC would complement rather than replace studies that will be carried out at other facilities with higher and more predictable rates. Like other photoproduction projects discussed here, the EIC could offer unique insights into the internal structures of charmonium-like states, and offer a U.S.\ option for particle physicists interested in fundamental aspects of hadron spectroscopy. With this potential in mind, detector designs should aim at optimizing the detection of final states relevant to the spectroscopy program.

\subsection{The PANDA Experiment}
\label{sec:panda}

The future PANDA (antiProton ANnihilations at DArmstadt) experiment, to be located at the Facility for Antiproton and Ion Research (FAIR) in Darmstadt, Germany, provides unique opportunities for studies of hadron spectroscopy in the high-intensity, lower-energy regime.  Its distinctive feature is the use of a beam of antiprotons, with a projected luminosity of $10^{31}$~cm$^{-2}$~s$^{-1}$ in PANDA's Phase One~\cite{PANDA:2021ozp}, which annihilate with protons in fixed targets.  The total integrated luminosity for Phase One
is expected to be about 0.5~fb$^{-1}$.

Being a strong-interaction process, $p\bar p$ annihilation features cross sections several orders of magnitude larger than those in (electromagnetic) $e^+ e^-$ or photoproduction processes; thus, PANDA will generate especially large data sets for analysis.  In addition, $p\bar p$ (unlike $e^+ e^-$) annihilation produces final states with no severe limitations on possible spin and parity quantum numbers, thus offering access to numerous exclusive channels.  Unlike $pp$ collisions, $p\bar p$ can produce single resonances $X$ via formation $p\bar p \! \to \! X$, {\it i.e.}, without the complication of additional hadrons in the final state.  With good momentum resolution ($\Delta p / p \! \le \! 5 \! \times \! 10^{-5}$), PANDA will then be able to perform detailed scans across resonance lineshapes for resonances up to 5.5~GeV/c$^2$ in mass.  As noted elsewhere in this document ({\it e.g.}, in Sec.~\ref{sec:JPAC}), precise lineshape information provides crucial information about the substructure of the state (for example, true resonance {\it vs.}$\!$ virtual state, molecular {\it vs.}$\!$ compact components).  In addition, with few constraints on final-state quantum numbers, searches for spin-isospin partner states, particularly for the exotic candidates $XY\! Z$, become much more straightforward.

Moreover, $p\bar p$ also naturally produces gluon-rich environments, which are particularly valuable in searches for glueball and hybrid candidates.  These processes are complementary to those of the dedicated electro- or photoproduction experiments (Sec.~\ref{sec:Photo}).

\section{Production of Hadronic Bound States}
\label{sec:production}

Heavy-quark hadrons, the narrow quarkonium states in particular, have played an important role as probes for various production processes involving gluons, as the heavy quarks are not present as constituents in beam particles. Their production characteristics test our understanding of hadroproduction, and can also be used to determine gluon structure functions inside the colliding particles. These topics have been reviewed by the EF06 working group, and we refer the reader to their report
. Heavy-quark hadrons are also important as probes of the quark-gluon plasma, since they are affected by the medium they traverse before decoupling from the other products of the collisions, as discussed in the report of EF07 working group
. 

Different production mechanisms are also important for probing the internal structure of the produced hadrons. Absolute production rates in different collisions, as well as their dependence on hadron
polarizations, production kinematical variables ({\it e.g.}, photon virtuality in photoproduction: see Sec.~\ref{sec:Photo}), or quantities probing the surrounding environment ({\it e.g.}, the multiplicity of the event: see Sec.~\ref{sec:lhcb}) can reveal the internal structure of exotic-hadron candidates, whose nature is often under dispute.
When studying such effects, comparison to the production of conventional hadrons is of key importance. We have touched on these subjects in various sections in this report. From the perspective of hadron spectroscopy, it is important to detect exotic hadrons in as many environments as possible, and so every experiment above the relevant production threshold should aim to look for new hadronic states, even though this may not be the main focus of the program.  Negative searches with quantitative upper limits are also important and should be published. 

\section{Theoretical Landscape and Prospects}

\subsection{Phenomenological Approaches}
\label{sec:Models}

As noted in the Introduction (Sec.~\ref{sec:introduction}), the possibility of multiquark exotic hadrons was anticipated even in the foundational papers of the quark model~\cite{Gell-Mann:1964ewy,Zweig:1964ruk,Zweig:1964jf}, using group-theoretical insights that (in modern language) amount to the conditions for forming a color singlet.  The prospect of multiquark states in the form of deuteron-like molecules of heavy-quark hadrons was first proposed~\cite{Voloshin:1976ap,DeRujula:1976zlg} almost immediately after the discovery of charm.  The 1970s also saw the first explicit calculations~\cite{Jaffe:1976ig,Jaffe:1976ih} of multiquark states in which all constituents reside in a single confinement volume, focusing upon light scalar mesons such as $a_0$ and $f_0$ that exhibit peculiar properties; both 4-quark and diquark-antidiquark configurations were considered. The importance of diquarks in several aspects of hadron phenomenology was assessed in~\cite{Wilczek:2004im}. Since deriving detailed properties of unstable multiquark states directly from the QCD Lagrangian is not yet feasible, much of the theory of more complex hadrons is QCD-motivated phenomenology. 

Of course, the theory landscape expanded dramatically with the 2003 discovery by Belle~\cite{Choi:2003ue} of $X(3872)$, the first heavy exotic-hadron candidate.  This section, which summarizes the Snowmass white paper~\cite{Brambilla:2022ura}, discusses many of the theoretical advances obtained by researchers exploring a wide variety of approaches, and identifies a number of important directions for future investigations.  Contributions to certain theoretical areas, such as the results of several detailed amplitude analyses and of lattice simulations, are discussed in Secs.~\ref{sec:JPAC} and \ref{sec:Lattice}, respectively.

Most (but not all) of the modern theoretical studies on multiquark hadrons implicitly or explicitly consider states containing at least one heavy quark ($m_{c,b} \! \gg \! \Lambda_{\rm QCD}$); this fact reflects not just the bulk of recent experimental discoveries, but also that the presence of heavy quarks makes both conventional and exotic hadrons much easier to identify. In the case of exotics, the presence of both heavy and light quarks allows decay modes that unambiguously reveal their multiquark nature. For example, in $Z_c^+\! \to \! J/\psi \pi^+$ decays, the $J/\psi$ reveals a $c\bar{c}$ pair, and $\pi^+$ reveals a $u\bar{d}$ pair. Neither can pop up in the decay process, and thus both are present in the $Z_c^+$.  Exotics might (and indeed, almost certainly do) mix with conventional hadrons of the same $J^{PC}$, but can have decays that would be unnatural for the conventional state [{\it e.g.}, the isospin-violating $J/\psi \rho$ mode of the $1^{++}$ $X(3872)$ {\it vs.} $\!$expectations for the conventional $\chi_{c1}(2P)$].  The large mass of the heavy quarks greatly reduces their kinetic energy, making it easier for them to form multiquark clusters with the light quarks, and the question of precisely how the quarks are organized within the full state remains to this day one of the key unresolved mysteries.  Some exotics lie rather close to di-hadron thresholds and have a natural molecular interpretation, while others do not.  It is quite possible that molecular, diquark, threshold-effect, and other interpretations hold for different exotics, or the states might even be superpositions of such configurations.  A closely related question is what other multiquark states should be expected: Do the $c$-containing and $b$-containing candidates follow parallel structures?  What kinds of multiplets do they fill, and what does this say about their substructure?  Tetraquark and pentaquark candidates have been observed; are heavy-quark hybrids and hexaquarks far behind?  Here we consider a number of interesting directions of research that have been carried out in recent years.  Clearly such a listing is merely indicative rather than exhaustive.

QCD sum rules, which exploit quark-hadron duality in two-point current correlators, can be used to identify exotic candidates if the current $J$ has the same quantum numbers as the physical state.  As an example, in the $s\bar s s \bar s$ sector, Ref.~\cite{Chen:2018kuu} obtains a pair of such states, one of which can be identified with $Y(2175)$, and the other with the recently observed $X(2400)$ in $e^+ e^- \! \to \! \phi \pi^+ \pi^-$ at BESIII~\cite{BESIII:2021lho}.

Molecular and diquark models are not the only ones used for multiquark states.  In addition, one may consider {\it hadroquarkonium}~\cite{Dubynskiy:2008mq}, in which the heavy $Q\bar Q$ pair forms a small core about which the light-quark cloud forms.  Alternately, exotics near di-hadron thresholds may be bound through quantum field-theory effects (instead of by light-meson exchanges, as in traditional molecular models)~\cite{Dong:2020hxe}.  Included in this category are so-called {\it triangle singularities} (effects from loop diagrams in which the two near-threshold component particles and a third ``bachelor'' particle form a triangular loop diagram), which may be probed through the exotic decay lineshape to glean information on the nature of the state~\cite{Swanson:2014tra,Pilloni:2016obd,Guo:2019twa}.

The proximity of threshold to many of the exotic candidates leads to important studies best undertaken using the tools of scattering theory.  The well-known effective-range expansion provides information on the nature of such a state through its parameters, the scattering length $a_0$ and the effective range $r_0$.  Weinberg long ago derived a criterion~\cite{Weinberg:1965zz} using $a_0$, $r_0$, and the binding energy of the state to determine if it is primarily composite (like the deuteron) or compact.  This criterion has been used to great effect in arguing, for example, that $X(3872)$ and $T_{cc}^+$ are essentially molecular ($D^0 \bar D^{*0}$ and $D^0 D^{*+}$, respectively)~\cite{Baru:2021ldu}.  However, another group using the same tools~\cite{Esposito:2021vhu} has argued that $X(3872)$ must have a substantial compact component.

In instances where the presence of a di-hadron threshold is important, scattering into more than one distinct channel can be crucial to the structure of the state, in which case one must develop a {\it coupled-channel} formalism to include all components.  One example of studies of this type for exotics~\cite{Takizawa:2012hy} treats $X(3872)$ as the combined effect of $D^0 \bar D^{*0}$, $D^+ \bar D^{*-}$, and $\chi_{c1}(2P)$ coupled components.  Of course, many other works include coupled-channel effects, and incorporating such improvements will likely be essential in any final theory of exotics.

Not only tetraquarks, but also pentaquarks like the $P_c$ states observed by LHCb~\cite{LHCb:2019kea} (and predicted long before, {\it e.g.}, in Ref.~\cite{Wu:2010jy}) have been studied in a variety of works.  For example, if $P_c(4457)$ is an $I \! = \! \frac 1 2$ $\Sigma_c \bar D^*$ molecular state, Ref.~\cite{Guo:2019fdo} argues that it will then have a large branching fraction to the isospin-breaking final state $J/\psi \Delta$ via rescattering effects.  Pentaquarks also naturally arise in models with diquarks as color-triplet diquark-triquark compounds, and as such are predicted to fill complete multiplets~\cite{Lebed:2017min,Ferretti:2020ewe}.

Indeed, the fact that SU(3)$_{\rm flavor}$ multiplets differ for molecular models (in which the allowed states' overall flavor quantum numbers are the combination of those from the component hadrons) and for compact or diquark models (in which the allowed states' overall flavor quantum numbers are determined by combining those of all the light quarks) has been noted by several authors ({\it e.g.}, Refs.~\cite{Ferretti:2020ewe,Maiani:2021tri}) as a crucial piece of information in discerning the nature of the states.  Heavy-quark spin symmetry also provides a valuable tool for revealing the character of the exotic states, both through their multiplet structure and their preferred decay modes~\cite{Cleven:2015era,Hosaka:2016ypm}.  For instance, some hidden-charm exotics preferentially decay to $J/\psi$ ($s_{c\bar c} \! = \! 1$) while others prefer $h_c$ ($s_{c\bar c} \! = \! 0$), which gives an important clue to their internal design.

The heaviness of quarks in the exotic states supplies a further valuable limit in which to perform computations: To lowest order in $\Lambda_{\rm QCD}/m_Q$, they act as static sources that permit the application of the Born-Oppenheimer (BO) approximation.  Using this starting point, one can build a systematic effective-field theory\footnote{Effective-field theories appear prominently in multiple areas relevant to the Snowmass Rare Processes Frontier, such as studies of dark matter, lepton-flavor violation, weak decays, and others.} in inverse powers of $\Lambda_{\rm QCD}/m_Q$~\cite{Braaten:2014qka,Brambilla:2017uyf}, which in turn can be used to identify the most physically significant operators contributing to the state structure.  The BO approximation has been applied phenomenologically as well, {\it e.g.}, in the dynamical diquark model~\cite{Giron:2019cfc}.

Diquark models, as suggested above, tend to fill larger spin (and isospin) multiplets due to the combinatorics of the large number of available diquark-component quantum numbers.  For instance, Ref.~\cite{Bedolla:2019zwg} predicts a very rich spectrum of all-heavy tetraquarks (18 in the $c\bar c c\bar c$ sector with $C \! = \! +$ between the di-$J/\psi$ threshold and 7380~MeV).  In comparison, the subsequent measurement of the di-$J/\psi$ spectrum in this range~\cite{LHCb:2020bwg} reveals two or three structures, including the resonance $X(6900)$.  This comparison can be applied to achieve progress in both directions: by seeking refined resolution of the experimental results that may uncover many more states, or by refining the model to eliminate artifacts that induce an excess of predicted states.

The consideration of diquarks as quasiparticles can also be used to make predictions for sectors not yet explored experimentally.  For example, a simple {\it ansatz\/} relating the $cc$ color antitriplet to the $c\bar c$ color singlet led to an accurate prediction for the $\Xi_{cc}^{++}$ mass~\cite{Karliner:2014gca}, and then later for $T_{cc}^+$~\cite{Karliner:2017qjm}, both of which were subsequently observed by LHCb~\cite{LHCb:2017iph,LHCb:2021auc}.

Double- (or multiply)-heavy (as opposed to hidden-heavy) exotics such as $T_{cc}^+$ are especially interesting because the attraction between two heavy quarks grows as $\,\alpha_s^2 m_Q$.  Thus, they may have strong-decay stability properties that states like $X(3872)$ lack, an important effect noted long ago for all-heavy systems~\cite{Ader:1981db}.  The $cc \bar u \bar d$ state was predicted in a molecular model to be just barely bound~\cite{Janc:2004qn}, long before the discovery of $T_{cc}^+$ by LHCb~\cite{LHCb:2021auc}, while the corresponding $T_{bb}^-$ state is expected to be strongly bound~\cite{Janc:2004qn,Karliner:2017qjm,Eichten:2017ffp}.  Indeed, for $T_{bb}^-$, separate molecular and compact states are possible~\cite{Meng:2020knc}, and $T_{bc}^0$ may present yet a third mixture~\cite{Karliner:2017qjm}.
Lattice simulations (discussed in Sec.~\ref{sec:Lattice}) predict bound $bb \bar u \bar d$ and $bb \bar u \bar s$ states.  Contemporary models study not just these bound states but their excited resonances as well~\cite{Meng:2021yjr}. 

The theoretical study of multiquark exotic hadrons remains tightly coupled to ongoing experimental advances.  Since no single paradigm has of yet explained all of the new states, each viable alternative must continue to be explored.  Strong discriminants to discern the structure of exotic states include uncovering their spin and flavor multiplet structure, investigating all distinct flavor sectors and comparing the states appearing in each one, examining patterns in their production and decay channels, and studying their detailed decay lineshapes.  Such studies also inform lattice-QCD efforts to identify and calculate the most incisive observables for discerning hadronic structure.  Dedicated effort to create a consortium of U.S.\ theorists for the purpose of enlarging individual research perspectives, potentially forming new collaborations, and training junior researchers, is strongly recommended.

\subsection{Amplitude Analysis}
\label{sec:JPAC}
All hadronic models, such as those discussed in Sec.~\ref{sec:Models}, manifest a particular set of assumptions about the dominant dynamical degrees of freedom and their interactions within the full hadronic state.  In this sense, models comprise a ``top-down'' strategy for studying strong-interaction physics. A quite fruitful alternate approach pursues a ``bottom-up'' scheme of imposing only the most fundamental properties of any quantum field theory: unitarity (probability conservation), analyticity of $S$-matrix amplitudes in terms of kinematical Lorentz invariants (a consequence of causality), and crossing symmetry (a property of all well-defined relativistic quantum theories).  Each of these features is actually more general than QCD itself, and together they provide strong constraints even in complicated regions of overlapping, broad hadronic resonances, coupled-channel rescatterings, and states formed from exotic combinations of degrees of freedom (glueballs, hybrids, and multiquark states).  On one hand, enforcing these principles provides a powerful method for analyzing complicated data sets while imposing a minimum of theoretical bias when interpreting the nature of their observed features; on the other hand, these principles place considerable constraints upon any model (or even the results of a lattice-QCD [LQCD] simulation) that is required to obey them.

While much of the basic formalism for such studies ({\it e.g.}, dispersion relations, lineshape analysis, Regge phenomenology) has existed since the 1960s, fully exploiting the large and complicated data sets that are being collected at modern experiments by employing state-of-the-art analytical and computational methods requires the collective efforts of not only multiple theorists, but experimentalists as well.  The Joint Physics Analysis Center (JPAC), a multi-institutional and international effort, precisely embodies this type of collaboration; their Snowmass white paper~\cite{JPAC:2022ipt} summarized here presents a general discussion of the application of their techniques to the study of hadronic systems, focusing upon their work related to hadronic spectroscopy and prospects for developing and applying novel analysis procedures.

The JPAC analysis methods have been applied to address important questions related to light exotic hadrons, like glueballs ({\it e.g}, distinguishing the scalar $f_0(1710)$ as possessing a strong glueball component~\cite{Rodas:2021tyb}) or hybrids ({\it e.g.}, identifying the peak $\pi_1(1400)$ as being an artifact of the true $J^{PC} \! = \! 1^{-+}$ $\pi_1(1600)$~\cite{JPAC:2017dbi,JPAC:2018zyd}). 
The heavy-quark sector presents its own unique complications in terms of identifying the nature of an observed state.  While the lowest quarkonium states are certainly sufficiently narrow to treat as isolated Breit-Wigner resonances, the exotic candidates like $X(3872)$, $Z_c(3900)$, $P_c(4312)$, {\it etc.}\ are far more complicated.  Many of these states lie quite close to di-hadron thresholds, possibly suggesting a molecular state structure, but definitely indicating the influence of the corresponding Fock component in their production and decay properties.  In such cases, the decay lineshape can provide crucial information.  For example, the pentaquark candidate $P_c(4312)$ seen at LHCb~\cite{LHCb:2019kea} lies only $\sim$~5~MeV below the $\Sigma^+_c \bar D^0$ threshold.  Depending upon which region of the complex Riemann surface for the reaction amplitude contains the pole corresponding to the state, its nature, and hence its effect upon the observed lineshape, can be quite different.  For instance, a below-threshold pole in the complex-energy plane on an unphysical Riemann sheet is called a {\it virtual state}, and it manifests as an attractive (but not bound) di-hadron state with a large scattering length and a distinctive threshold-cusp lineshape.  The JPAC analysis~\cite{Fernandez-Ramirez:2019koa} of LHCb data shows that $P_c(4312)$ is more likely a virtual than a true bound state of $\Sigma^+_c \bar D^0$.  A similar study of $Z_c(3900)$ finds that not enough detail appears in the current data allowing one to draw an analogous conclusion~\cite{Albaladejo:2015lob,Pilloni:2016obd}; these theory analyses can thus point the way toward specific future experimental measurements at sufficient precision to resolve the issue.

Many light hadrons, as well as many of the exotic-hadron candidates, possess prominent 3-body decays.  Since $3 \! \to \! 3$ processes have 8 Mandelstam invariants, the landscape of energy-dependent functions required to satisfy the core $S$-matrix requirements is much more extensive than for $2 \! \to \! 2$ processes (2 invariants), and both the GWU (George Washington U.) and JPAC groups have contributed to the development of formalism for these complicated reactions~\cite{Mai:2017vot,Jackura:2018xnx,Mikhasenko:2019vhk}.  In the light sector, reactions like $\gamma^* \! \to \! 3\pi$ are essential for understanding hadron vacuum-polarization contributions to the muon anomalous magnetic moment $(g-2)_\mu$, and the $\gamma^* \! \to \! 3\pi$ amplitude is closely related to the process $\omega \! \to \! 3\pi$ studied by JPAC~\cite{JPAC:2020umo}.  In the heavy sector, the newly discovered double-charm state $T^+_{cc}$~\cite{LHCb:2021auc} is so far seen only as a peak in the 3-body channel $D^0 D^0 \pi^+$, meaning that $3 \! \to \! 3$ amplitude techniques need further development to permit full analysis of such states in the future.

The method of applying fundamental quantum field-theory principles to the detailed study of strong-interaction amplitudes has a long history, but after decades of falling into disuse, its rebirth in modern applications has found great utility in answering physical questions that cannot easily be addressed in other ways.  Experiments at both existing facilities (LHC, BESIII, JLab, Belle~II), and upcoming ones (EIC, PANDA) will provide enormous data sets that will be best analyzed using these techniques.  Since the relevant calculations require not only a great deal of intricate work and advanced computing power, not to mention insight into the detailed capabilities of experiments, they are best carried out by a tight collaboration of theorists and experimentalists.  Furthermore, modern LQCD efforts (Sec.~\ref{sec:Lattice}) are partially driven to decrease the pion mass or increase the number of decay channels. As such, they require an advanced amplitude framework to produce successful results.  JPAC provides a working model for this collaboration. It also promotes an effective bridge between the nuclear and high-energy communities by their common interest in hadron spectroscopy. 
Such activities ought to be supported.

\subsection{Lattice QCD}
\label{sec:Lattice}
Lattice Quantum Chromodynamics (LQCD) simulates the full theory of QCD by mapping its path integral onto a discretized grid of finite volume in Euclidean spacetime, thus allowing for its numerical evaluation.  LQCD provides one of the very few known rigorous approaches for performing first-principles calculations in strong-interaction physics, particularly in the nonperturbative regime. The uncertainties in its calculations are (at least in principle) all quantifiable and systematically improvable; indeed, the dependence of its results on finite volume or on quark masses can themselves be used to discern interesting physical phenomena.  The interplay between LQCD uncertainties can be quite subtle and quantity-dependent, and can only be disentangled through careful theoretical analysis.  And of course, LQCD simulations are ultimately limited by computing power in achieving suitable signal-to-noise ratios for producing meaningful numerical results.  Nevertheless, LQCD simulations not only present a unique framework to support and complement experimental searches, but also provide both constraints and enhancements for phenomenologically viable models of strong-interaction physics.  The application of LQCD to hadronic spectroscopy and its future prospects have been recently reviewed for the Snowmass Exercise in Ref.~\cite{Bulava:2022ovd}.

The last decade has seen the precision and predictive power of LQCD hadron spectroscopy dramatically improve, especially for bound states that can be treated as stable under strong decay.  In such cases, masses have been determined at $< \! 5\%$ precision, with all systematic uncertainties being taken into account.  The most significant advances, however, have been achieved in calculations of states near (below or above) strong-decay thresholds, particularly using the {\it L\"uscher formalism}~\cite{Luscher:1986pf,Luscher:1990ux} for extracting scattering amplitudes from the finite-volume dependence of energies.  Since nearly every exotic-hadron candidate is expected to possess strong decay modes, such techniques are absolutely crucial for their study.  Improvements in both algorithms and analytic understanding have begun to impact lattice calculations of
conventional and exotic light- and heavy-quark hadronic resonances, and will play an important role in the future development of LQCD hadron spectroscopy.

In this Section we present a (very brief and selective) overview of relevant developments in LQCD studies of hadron spectroscopy and closely related phenomena, and for each category we illustrate the type of advances that may be expected within the next several years.

(1) The simplest possible strong-decay process is one with only two spinless mesons in the final state.  To this end, the modes $\rho\rightarrow\pi\pi$ and $K^\ast\rightarrow K\pi$ have been studied by several lattice groups, and provide proof of concept for the reliability of the analysis methods. In fact, LQCD simulations using the physical value of $m_\pi$ are now possible; but they are quite costly, not only due to increasing noise in the results, but also because many more multipion decay channels open up, which leads to added complications in the extraction of resonance parameters.  The $\rho$ and $K^\ast$ are sufficiently narrow to be treated as Breit-Wigner resonances, while a few pioneering studies have extracted the much broader scalars $\sigma \! \to \! \pi \pi$~\cite{Briceno:2016mjc,Guo:2018zss}, $\kappa \! \to \! K \pi$~\cite{Brett:2018jqw,Rendon:2020rtw,Wilson:2019wfr}, and $D^*_0 \! \to \! D \pi$~\cite{Gayer:2021xzv,Moir:2016srx,Mohler:2012na}  Even so, one may anticipate the light-quark mass dependence of all of these processes to be mapped out in more detail within the next five years (including more rigorous determinations of their parameters, using the amplitude techniques described in Sec.~\ref{sec:JPAC}), thereby providing valuable information on resonance structure, as well as calculations that are directly relevant to experiment.

(2) In most 2-body strong decays (and in particular for most observed exotic states), more than one distinct decay channel occurs, and/or the resulting hadrons carry nonzero spin ({\it e.g.}, $X(3872) \! \to \bar D^0 D^{\ast 0}$, $J/\psi \, \omega$, {\it etc.}).  Managing the resulting coupled-channel rescattering leads to more complicated matrix-valued scattering amplitudes (increasing in difficulty with the number of channels) that must be disentangled in order to extract resonance parameters.  Spin leads to further complications, not only due to the appearance of multiple individual state components, but also because many more partial waves (which often mix) can arise for nonzero spin.  While several pioneering LQCD calculations have been performed in the light-quark sector~\cite{Dudek:2014qha,Briceno:2017qmb,Woss:2019hse}, the only coupled-channel study to date in the heavy-quark sector examines $D\pi$-$D\eta$-$D_s\bar{K}$~\cite{Moir:2016srx} and $D\bar D$-$D_s\bar D_s$~\cite{Prelovsek:2020eiw} mixing.  A more complete investigation of both exotic and conventional resonance scattering matrices, including coupled-channel (at least up to three states) and nonzero-spin effects, is a central goal for LQCD in the next five years.
   

(3) Studies of the spectroscopic properties and structure of exotic hadrons that are simultaneously accessible to LQCD and to experiment are particularly valuable.  On one hand, experimental results can provide insight into the limitations of LQCD calculations, while on the other hand LQCD can predict the existence of new states and decay modes, to provide impetus for further experimental studies.  LQCD simulations also develop physical insight by allowing systems to be studied with varying quark masses.  Also, some kinematical regions/quantum numbers/decay modes might be more amenable to examination in one approach than in the other.  While LQCD can reliably probe regions below or slightly above the lowest strong-decay threshold, only a subset of the observed exotic-hadron candidates appears in such energy intervals.  As examples of this dual approach, bound states of flavor content $bb\bar u\bar d$ and $bb\bar u\bar s$ are firmly established by LQCD~\cite{Bicudo:2015vta,Francis:2016hui,Hudspith:2020tdf,Leskovec:2019ioa,Junnarkar:2018twb}, but producing sufficient numbers of $bb$ (rather than $b\bar b$) states for experimental discovery will pose a formidable challenge. LQCD calculations also predict a supermultiplet of hybrid states, above strong-decay thresholds, in both charmonium~\cite{HadronSpectrum:2012gic} and bottomonium~\cite{Ryan:2020iog}. The experimentally observed states $X(3872)$ ($c\bar c q\bar q$) and $T_{cc}^+$ ($cc\bar u\bar d$) lie very near open-charm thresholds, and while LQCD calculations will contribute valuable insights, the extreme proximity of these examples to open-flavor thresholds limits the precision in the calculation of their detailed properties.  Current LQCD results are reliable to $O(10~{\rm MeV})$~\cite{Prelovsek:2013cra,Padmanath:2022cvl}, while a few MeV is the goal. Including isospin dependence is one path to aspiring for this accuracy; while LQCD studies that incorporate isospin breaking are in their infancy, they are worth pursuing in the next 10 years.


(4) LQCD studies of baryon systems provide not only the gateway to understanding light nuclei, but also calculate matrix elements (which serve as inputs to nuclear effective-field theories) that can be used to eliminate many systematic uncertainties in beyond-Standard Model (BSM) searches.  A reliable, robust prediction of the spectra and matrix elements of two-nucleon systems, with careful control over systematic uncertainties, is an ambitious but achievable goal for LQCD in the next decade.  In addition, LQCD calculations have predicted the existence~\cite{Green:2021qol} (at $m_\pi \! = \! 420$~MeV) of an $H$-dibaryon, a $udsuds$ state below the $\Lambda \Lambda$ threshold that, if confirmed, would represent the first {\it hexaquark\/} state.


(5) The LQCD community also faces longer-term challenges in the calculation of decays involving three or more particles in the final state.  This problem has received a substantial investment of effort in recent years. For example, while several LQCD simulations have considered $3\pi$ scattering, only one exploratory study thus far has considered a resonance that strongly decays to three hadrons~\cite{Mai:2021nul}. Nevertheless, the structure of scattering amplitudes with three or more final hadrons can be quite intricate, and substantial future theoretical work will be needed to provide robust tools that can extract the desired resonance information from simulations.


(6) LQCD simulations to calculate resonance electromagnetic and weak transitions have commenced only recently, but show great potential for producing novel types of results.  The pioneering work on $\pi\gamma \! \rightarrow \! \rho \! \to \! \pi\pi$~\cite{Briceno:2015dca,Alexandrou:2018jbt} opens the way to simulations of $N\gamma\rightarrow\Delta\rightarrow N\pi$ and $K\gamma\rightarrow K^\ast\rightarrow K\pi$.  The same techniques can be applied in the medium term to weak-current transitions between stable states and resonances, or even resonance-to-resonance transitions, thereby developing an incisive probe into the structure of both conventional and exotic hadrons.  Such calculations are also timely and relevant due to their significance for BSM searches as in, {\it e.g.}, $B\rightarrow D^\ast \ell\bar{\nu}$; $D^\ast\rightarrow D\pi$, a decay that is prominent in current studies of lepton-flavor-universality violation.  In each of these cases, further results can be expected within five years.
%

The essential resources necessary to guarantee the advances outlined above and to maintain the impact of LQCD hadronic physics in the future include continued access to ever-improving computing resources.  Of equal importance is to sustain investment in human capital, particularly through support for younger researchers.  LQCD funding in the U.S.\ in recent decades has largely flowed from the Department of Energy to the USQCD Collaboration (which represents almost all U.S.\ lattice groups), and we strongly recommend such continued support.  In addition, we recommend fostering partnerships with the DOE's SciDAC (Scientific Discovery through Advanced Computation) program and other federal advanced-computing initiatives.

\section{Recommendations}

The explosion of new experimental findings in hadron spectroscopy over the past two decades has been so extensive and unexpected that the theoretical community still grapples with very fundamental unanswered questions: Do compact tetraquarks or pentaquarks exist?  What is their dominant color-binding scheme (diquarks, hadroquarkonia, \dots)?  Do hadron pairs create loosely bound deuteron-like  ``molecules"?  In which configurations?  Do gluons play the role of a valence hadron constituent (hybrids, glueballs)?  What is the level of mixing of various bound-state configurations?  Answering these questions must be a priority of particle physics in the coming years.  A better understanding of such phenomena is not only important in its own right, but will likely have a significant impact on quantifying hadronic uncertainties in searches for new physics, the modeling of hadronic matter in neutron stars, and even shaping the phenomenology of any strongly coupled extensions of the Standard Model. 

Fortunately, many ongoing experimental projects are likely to produce a wealth of new results that will provide insights to resolve these questions.  While hadron spectroscopy was often not among the original primary motivations for these experiments, the sheer number and quality of novel discoveries demand greater attention to the existing scientific opportunities.  Specifically, we advocate first for the support of funding agencies in order to carry out dedicated data analyses.  The relevant experimental facilities are either already operating now, or are expected to be running within the next decade, and therefore do not require a separate request.  Thus, we recommend primarily for support in developing human capital: the training and funding of young scientists to carry out this ambitious program.  Additionally, since discoveries of new types of hadronic phenomena often produce the most cited publications of an experimental collaboration [$X(3872)$ for Belle, $Z_c(3900)$ for BESIII, $P_c$ states for LHCb] as well as attract significant public attention, such investments easily pay off. 

LHCb Runs 1 \& 2 produced major discoveries in the form of the narrow pentaquark states, a variety of broader tetraquark states, open-charm tetraquark states, double-charmonium tetraquark structures, and a double-charm narrow tetraquark state, in addition to enriching the spectroscopy of conventional heavy baryons and mesons. The upcoming Runs 3 \& 4, with the upgraded LHCb detector being able to process the higher LHC instantaneous luminosity, is likely to produce an equally large number of new exciting results.  Overall, LHCb has the broadest potential to significantly advance hadron spectroscopy over the next two decades, by studying exotic states with a large variety of quark configurations.  
Continued U.S.\ support for this experiment is a high priority from the point of view of hadron spectroscopy.  Detector R\&D for LHCb Upgrade 2 (Run 5) is needed to exploit the full heavy-flavor physics potential of the luminosity delivered by the LHC\@. The upgrade of the LHCb EM calorimeter would be important not only for searches for BSM physics, but also for hadron spectroscopy. Installation of tracking chambers inside the magnet will also be important.

The Belle~II detector, operating at the highest-luminosity $e^+e^-$ collider (SuperKEKB) at the $B\bar{B}$ threshold, provides unique opportunities in heavy-hadron physics.  Scanning above the $\Upsilon(4S)$ resonance as high in energy as possible for exotic states with $b\bar{b}$ content (tetraquarks, hybrids?), including charged $Z_b$ states, is its most important distinctive window into exotic-hadron physics.  Exploiting initial-state radiation allows the exploration of equivalent structures with $c\bar{c}$ content, in competition with BESIII program. Two-photon collisions and double-$c\bar c$ processes offer additional unique windows into hadron spectroscopy for Belle~II.  Hadron spectroscopy will benefit from continued U.S.\ support for researchers in this program. 

The CMS and ATLAS experiments already run at the maximum luminosity delivered by the LHC\@. This fact alone constitutes a unique discovery potential in modes in which hadron identification is not crucial for background suppression, like double-quarkonium states decaying to a pair of dimuons.  The confirmation or even discovery of a number of conventional (and now, exotic) heavy-quark hadrons has already occurred, and may continue in the future.  Adjustments to the allocation of trigger bandwidth and of analysis resources to such channels is strongly recommended.

The BESIII program presents unique capabilities
in hadron spectroscopy via scanning $e^+e^-$ annihilation above the $D\bar{D}$ threshold, as well as via a very large sample accumulated at the $J/\psi$ resonance for light-hadron spectroscopy. This program has produced many important exotic-hadron candidates with hidden charm, especially the $J^{PC} \! = \! 1^{--}$ $Y$ states, as well as numerous findings for the conventional charmonium system and transitions among all of these states. Supporting the participation of U.S.\ physicists in the remaining years of this program (which includes an upgrade to energies that allow studies of charmed hyperons) not only leverages current scientific opportunities, but also positions such researchers to transition to highly visible roles at future Super Tau-Charm facilities. While such undertakings will be led by the other countries, supporting a degree of U.S.\ research participation in such programs leverages investments made by others to benefit the U.S.\ particle physics community. Similar comments apply to the future PANDA experiment, designed to scan $p\bar{p}$ annihilation near the $c\bar{c}$ threshold, and provide detailed lineshape information for hadronic states.

Hadronic spectroscopy will also be studied at U.S.\ experimental facilities, namely, the currently running GlueX Experiment at Jefferson Lab and the Electron-Ion Collider (EIC) expected to begin taking data in 2030.  In both cases, the primary production mechanism is photoproduction on proton and nuclear targets, which offers special insight into inner structure of the produced hadrons.  While one central original goal of GlueX is the study of light-quark hybrid mesons like $\pi_1(1600)$, the experiment will also probe threshold-region production of $P_c$ pentaquarks, as well as search for the hidden-strangeness analogues of $XYZP_c$ states. An energy upgrade of CEBAF would make such observations (and of the original hidden-charm $XYZ$ states) more likely.  
The EIC also has a good potential to photoproduce charmonium-like exotic hadrons, though the rates will be limited by the electromagnetic nature of the production cross sections.  
 Clearly, sustaining the support for researchers to perform these unique studies must be maintained.

In light of the fact that the latest discoveries in hadron physics no longer support a single simple paradigm to describe their spectroscopy or substructure, the theoretical landscape has become much more varied in its approaches.  Most of the center of mass for pursuing these studies no longer resides in the U.S., but dedicated efforts are being made  to bring together U.S.\ theorists for the purpose of joint projects and to support junior researchers ({\it e.g.}, a proposal to form a collaboration under the aegis of a Department of Energy Topical Collaboration in Nuclear Theory).  Of particular note is how such a proposal bridges the traditional particle/nuclear physics divide.  Such efforts, both collectively and individually, seamlessly incorporate the participation of non-U.S.\ researchers as well, and provide intellectual space for multiple research perspectives, including studies using threshold effects, diquark models, lattice-QCD studies, effective-field theories, and others.  Support for such initiatives forms a primary recommendation of this group.

The Joint Physics Analysis Center (JPAC) exhibits a model of a successful consortium of multiple researchers studying complex problems in hadronic analysis.  In this case, both theorists and experimentalists work together to analyze complicated data sets of overlapping resonances, thresholds, and other effects in hadronic data.  While the fundamental tools used by JPAC are cornerstone properties of $S$-matrix theory (unitarity, causality, crossing symmetry) that have been exploited for decades, the availability of multiple expert researchers and high-quality computing power means that these principles can be applied in new and sophisticated ways, such as by incorporating the results of lattice-QCD simulations, adopting advanced statistical approaches, and even applying machine-learning algorithms.  We strongly advocate for continued support of this type of intensive, collective effort to advance knowledge of hadronic physics.

We strongly recommend the continued support of lattice-QCD research not only through existing funding streams, but also recommend fostering relationships with other federal programs such as SciDAC whose mission includes modeling of complex systems.

\section*{Acknowledgements}
\noindent We gratefully acknowledge the hard work of the RF7 Subtopic Conveners, the authors of the contributed white papers, and all individuals who attended the RF7 workshops.  RL is supported by the US National Science Foundation (NSF) under Grants No. PHY-18030912 and PHY-2110278.  TS is supported by NSF under Grant No. PHY-2102879.

\clearpage

\addcontentsline{toc}{section}{References}
\bibliographystyle{JHEP}
\bibliography{main}

\providecommand{\href}[2]{#2}\begingroup\raggedright\begin{thebibliography}{100}

\bibitem{Gell-Mann:1964ewy}
M.~Gell-Mann, \emph{{A Schematic Model of Baryons and Mesons}},
  \href{https://doi.org/10.1016/S0031-9163(64)92001-3}{\emph{Phys.\ Lett.}
  {\bfseries {\bf 8}} (1964) 214}.

\bibitem{Zweig:1964ruk}
G.~Zweig, ``{\em An SU(3) Model for Strong Interaction Symmetry and Its
  Breaking. Version~1}.''
  \url{https://cds.cern.ch/record/352337/files/CERN-TH-401.pdf}, 1964.

\bibitem{Zweig:1964jf}
G.~Zweig, \emph{{An SU(3) Model for Strong Interaction Symmetry and Its
  Breaking. Version~2 {\rm (1964)}}},  in \emph{{Developments in the Quark
  Theory of Hadrons, Vol.\ 1: 1964--1978}}, D.~Lichtenberg and S.~Rosen, eds.,
  p.~22, Hadronic Press, Nonantum, MA (1980).

\bibitem{Choi:2003ue}
{\scshape Belle} collaboration, \emph{{Observation of a Narrow Charmonium-Like
  State in Exclusive $B^{\pm} \to K^{\pm} \pi^+ \pi^- J /\psi$ Decays}},
  \href{https://doi.org/10.1103/PhysRevLett.91.262001}{\emph{Phys.\ Rev.\
  Lett.} {\bfseries 91} (2003) 262001}
  [\href{https://arxiv.org/abs/hep-ex/0309032}{{\ttfamily hep-ex/0309032}}].

\bibitem{LHCbCollaboration:2806113}
{\scshape LHCb} collaboration, \emph{{Future Physics Potential of LHCb, {\rm
  LHCb-PUB-2022-012, CERN-LHCb-PUB-2022-012}}},  in \emph{{2022 Snowmass Summer
  Study}}, 2022,
  \href{https://cds.cern.ch/record/2806113}{https://cds.cern.ch/record/2806113}.

\bibitem{LHCb:2019kea}
{\scshape LHCb} collaboration, \emph{{Observation of a Narrow Pentaquark State,
  $P_c(4312)^+$, and of Two-Peak Structure of the $P_c(4450)^+$}},
  \href{https://doi.org/10.1103/PhysRevLett.122.222001}{\emph{Phys.\ Rev.\
  Lett.} {\bfseries {\bf 122}} (2019) 222001}
  [\href{https://arxiv.org/abs/1904.03947}{{\ttfamily 1904.03947}}].

\bibitem{LHCb:2020jpq}
{\scshape LHCb} collaboration, \emph{{Evidence of a $J/\psi\Lambda$ Structure
  and Observation of Excited $\Xi^-$ States in the $\Xi^-_b \to J/\psi\Lambda
  K^-$ Decay}}, \href{https://doi.org/10.1016/j.scib.2021.02.030}{\emph{Sci.\
  Bull.} {\bfseries {\bf 66}} (2021) 1278}
  [\href{https://arxiv.org/abs/2012.10380}{{\ttfamily 2012.10380}}].

\bibitem{LHCb-PAPER-2022-031}
{\scshape LHCb} collaboration, ``{\em Observation of a $J/\psi\Lambda$
  Resonance Consistent with a Strange Pentaquark Candidate in $B^-\to
  J/\psi\Lambda\bar p$ Decays}.'' \url{https://indico.cern.ch/event/1176505/}
  (in preparation), 2022.

\bibitem{LHCb:2021chn}
{\scshape LHCb} collaboration, \emph{{Evidence for a New Structure in the
  $J/\psi p$ and $J/\psi \bar{p}$ Systems in $B_s^0 \to J/\psi p \bar{p}$
  Decays}}, \href{https://doi.org/10.1103/PhysRevLett.128.062001}{\emph{Phys.\
  Rev.\ Lett.} {\bfseries {\bf 128}} (2022) 062001}
  [\href{https://arxiv.org/abs/2108.04720}{{\ttfamily 2108.04720}}].

\bibitem{LHCb:2013kgk}
{\scshape LHCb} collaboration, \emph{{Determination of the $X(3872)$ Meson
  Quantum Numbers}},
  \href{https://doi.org/10.1103/PhysRevLett.110.222001}{\emph{Phys.\ Rev.\
  Lett.} {\bfseries {\bf 110}} (2013) 222001}
  [\href{https://arxiv.org/abs/1302.6269}{{\ttfamily 1302.6269}}].

\bibitem{LHCb:2015jfc}
{\scshape LHCb} collaboration, \emph{{Quantum Numbers of the $X(3872)$ State
  and Orbital Angular Momentum in Its $\rho^0 J/\psi$ decay}},
  \href{https://doi.org/10.1103/PhysRevD.92.011102}{\emph{Phys.\ Rev.\ D}
  {\bfseries {\bf 92}} (2015) 011102}
  [\href{https://arxiv.org/abs/1504.06339}{{\ttfamily 1504.06339}}].

\bibitem{LHCb:2020xds}
{\scshape LHCb} collaboration, \emph{{Study of the Lineshape of the
  $\chi_{c1}(3872)$ State}},
  \href{https://doi.org/10.1103/PhysRevD.102.092005}{\emph{Phys.\ Rev.\ D}
  {\bfseries {\bf 102}} (2020) 092005}
  [\href{https://arxiv.org/abs/2005.13419}{{\ttfamily 2005.13419}}].

\bibitem{LHCb:2020fvo}
{\scshape LHCb} collaboration, \emph{{Study of the $\psi_2(3823)$ and
  $\chi_{c1}(3872)$ States in $B^+ \rightarrow \left(
  J/\psi\pi^+\pi^-\right)K^+$ Decays}},
  \href{https://doi.org/10.1007/JHEP08(2020)123}{\emph{JHEP} {\bfseries {\bf
  08}} (2020) 123} [\href{https://arxiv.org/abs/2005.13422}{{\ttfamily
  2005.13422}}].

\bibitem{LHCb:2022bly}
{\scshape LHCb} collaboration, \emph{{Observation of Sizeable $\omega$
  Contribution to $\chi_{c1}(3872)\to\pi^+\pi^-J/\psi$ Decays}},
  \href{https://arxiv.org/abs/2204.12597}{{\ttfamily 2204.12597}}.

\bibitem{Belle:2011wdj}
{\scshape Belle} collaboration, \emph{{Observation of $X(3872)\to J/\psi
  \gamma$ and Search for $X(3872)\to\psi'\gamma$ in B Decays}},
  \href{https://doi.org/10.1103/PhysRevLett.107.091803}{\emph{Phys.\ Rev.\
  Lett.} {\bfseries {\bf 107}} (2011) 091803}
  [\href{https://arxiv.org/abs/1105.0177}{{\ttfamily 1105.0177}}].

\bibitem{LHCb:2014jvf}
{\scshape LHCb} collaboration, \emph{{Evidence for the Decay
  $X(3872)\rightarrow\psi(2S)\gamma$}},
  \href{https://doi.org/10.1016/j.nuclphysb.2014.06.011}{\emph{Nucl.\ Phys.\ B}
  {\bfseries {\bf 886}} (2014) 665}
  [\href{https://arxiv.org/abs/1404.0275}{{\ttfamily 1404.0275}}].

\bibitem{LHCb:2020bwg}
{\scshape LHCb} collaboration, \emph{{Observation of Structure in the $J
  /\psi$-Pair Mass Spectrum}},
  \href{https://doi.org/10.1016/j.scib.2020.08.032}{\emph{Sci.\ Bull.}
  {\bfseries {\bf 65}} (2020) 1983}
  [\href{https://arxiv.org/abs/2006.16957}{{\ttfamily 2006.16957}}].

\bibitem{LHCb:2021auc}
{\scshape LHCb} collaboration, \emph{{Study of the Doubly Charmed Tetraquark
  $T_{cc}^+$}}, \href{https://doi.org/10.1038/s41467-022-30206-w}{\emph{Nature
  Commun.} {\bfseries {\bf 13}} (2022) 3351}
  [\href{https://arxiv.org/abs/2109.01056}{{\ttfamily 2109.01056}}].

\bibitem{LHCb:2020bls}
{\scshape LHCb} collaboration, \emph{{A Model-Independent Study of Resonant
  Structure in $B^+\to D^+D^-K^+$ Decays}},
  \href{https://doi.org/10.1103/PhysRevLett.125.242001}{\emph{Phys.\ Rev.\
  Lett.} {\bfseries {\bf 125}} (2020) 242001}
  [\href{https://arxiv.org/abs/2009.00025}{{\ttfamily 2009.00025}}].

\bibitem{LHCb:2020pxc}
{\scshape LHCb} collaboration, \emph{{Amplitude Analysis of the $B^+\to
  D^+D^-K^+$ Decay}},
  \href{https://doi.org/10.1103/PhysRevD.102.112003}{\emph{Phys.\ Rev.\ D}
  {\bfseries {\bf 102}} (2020) 112003}
  [\href{https://arxiv.org/abs/2009.00026}{{\ttfamily 2009.00026}}].

\bibitem{LHCb-PAPER-2022-026}
{\scshape LHCb} collaboration, ``{\em First Observation of a Doubly Charged
  Tetraquark and Its Neutral Partner}.''
  \url{https://indico.cern.ch/event/1176505/} (in preparation), 2022.

\bibitem{LHCb:2017uwr}
{\scshape LHCb} collaboration, \emph{{Observation of Five New Narrow
  $\Omega_c^0$ States Decaying to $\Xi_c^+ K^-$}},
  \href{https://doi.org/10.1103/PhysRevLett.118.182001}{\emph{Phys.\ Rev.\
  Lett.} {\bfseries {\bf 118}} (2017) 182001}
  [\href{https://arxiv.org/abs/1703.04639}{{\ttfamily 1703.04639}}].

\bibitem{LHCb:2020tqd}
{\scshape LHCb} collaboration, \emph{{First Observation of Excited $\Omega_b^-$
  States}}, \href{https://doi.org/10.1103/PhysRevLett.124.082002}{\emph{Phys.\
  Rev.\ Lett.} {\bfseries {\bf 124}} (2020) 082002}
  [\href{https://arxiv.org/abs/2001.00851}{{\ttfamily 2001.00851}}].

\bibitem{LHCb:2017iph}
{\scshape LHCb} collaboration, \emph{{Observation of the Doubly Charmed Baryon
  $\Xi_{cc}^{++}$}},
  \href{https://doi.org/10.1103/PhysRevLett.119.112001}{\emph{Phys.\ Rev.\
  Lett.} {\bfseries {\bf 119}} (2017) 112001}
  [\href{https://arxiv.org/abs/1707.01621}{{\ttfamily 1707.01621}}].

\bibitem{LHCb:2020sey}
{\scshape LHCb} collaboration, \emph{{Observation of Multiplicity Dependent
  Prompt $\chi_{c1}(3872)$ and $\psi(2S)$ Production in $pp$ Collisions}},
  \href{https://doi.org/10.1103/PhysRevLett.126.092001}{\emph{Phys.\ Rev.\
  Lett.} {\bfseries {\bf 126}} (2021) 092001}
  [\href{https://arxiv.org/abs/2009.06619}{{\ttfamily 2009.06619}}].

\bibitem{LHCbCollaboration:2776420}
{\scshape LHCb} collaboration, ``{\em Framework TDR for the LHCb Upgrade II:
  Opportunities in Flavour Physics, and Beyond, in the HL-LHC Era}.''
  \url{https://cds.cern.ch/record/2776420}, 2021.

\bibitem{Chistov:2022rht}
R.~Chistov, V.~Papadimitriou, S.~Polikarpov, A.~Pompili and
  A.~Sanchez-Hernandez, \emph{{Review of CMS Contribution to Hadron
  Spectroscopy (Snowmass 2021 White Paper)}},
  \href{https://arxiv.org/abs/2204.06667}{{\ttfamily 2204.06667}}.

\bibitem{CMS:2016liw}
{\scshape CMS} collaboration, \emph{{Observation of $\Upsilon(1S)$ Pair
  Production in Proton-Proton Collisions at $ \sqrt{s}=8 $~TeV}},
  \href{https://doi.org/10.1007/JHEP05(2017)013}{\emph{JHEP} {\bfseries {\bf
  05}} (2017) 013} [\href{https://arxiv.org/abs/1610.07095}{{\ttfamily
  1610.07095}}].

\bibitem{CMS:2020qwa}
{\scshape CMS} collaboration, \emph{{Measurement of the $\Upsilon(1S)$ Pair
  Production Cross Section and Search for Resonances Decaying to
  $\Upsilon(1S)\mu^+\mu^-$ in Proton-Proton Collisions at $\sqrt{s} =$
  13~TeV}}, \href{https://doi.org/10.1016/j.physletb.2020.135578}{\emph{Phys.\
  Lett.\ B} {\bfseries {\bf 808}} (2020) 135578}
  [\href{https://arxiv.org/abs/2002.06393}{{\ttfamily 2002.06393}}].

\bibitem{LHCb:2018uwm}
{\scshape LHCb} collaboration, \emph{{Search for Beautiful Tetraquarks in the
  $\Upsilon(1S)\mu^+\mu^-$ Invariant-Mass Spectrum}},
  \href{https://doi.org/10.1007/JHEP10(2018)086}{\emph{JHEP} {\bfseries {\bf
  10}} (2018) 086} [\href{https://arxiv.org/abs/1806.09707}{{\ttfamily
  1806.09707}}].

\bibitem{CMS-PAS-BPH-21-003}
{\scshape CMS} collaboration, ``{\em Observation of New Structures in the
  $J/\psi J/\psi$ Mass Spectrum in $pp$ Collisions at $\sqrt{s}$=13~TeV}.''
  \url{https://cds.cern.ch/record/2815336/}, 2022.

\bibitem{ATLAS-CONF-2022-040}
{\scshape ATLAS} collaboration, ``{\em Observation of an Excess of
  Di-Charmonium Events in the Four-Muon Final State with the ATLAS Detector}.''
  \url{http://cds.cern.ch/record/2815676}, 2022.

\bibitem{CMS:2019kbn}
{\scshape CMS} collaboration, \emph{{Study of the ${B}^{+}\to
  {J}/\psi\overline{\Lambda}{p}$ Decay in Proton-Proton Collisions at
  $\sqrt{s}$ = 8 TeV}},
  \href{https://doi.org/10.1007/JHEP12(2019)100}{\emph{JHEP} {\bfseries {\bf
  12}} (2019) 100} [\href{https://arxiv.org/abs/1907.05461}{{\ttfamily
  1907.05461}}].

\bibitem{ATLAS:2019keh}
{\scshape ATLAS} collaboration, ``{\em Study of $J/\psi \, p$ Resonances in the
  $\Lambda_b \to J/\psi \, p K^-$ Decays in $pp$ Collisions at $\sqrt{s}=7$ and
  8~TeV with the ATLAS Detector}.''
  \url{https://cds.cern.ch/record/2693957/files/ATLAS-CONF-2019-048.pdf}, 2019.

\bibitem{CMS:2013jru}
{\scshape CMS} collaboration, \emph{{Observation of a Peaking Structure in the
  $J/\psi \, \phi$ Mass Spectrum from $B^{\pm} \to J/\psi \phi K^{\pm}$
  Decays}}, \href{https://doi.org/10.1016/j.physletb.2014.05.055}{\emph{Phys.\
  Lett.\ B} {\bfseries {\bf 734}} (2014) 261}
  [\href{https://arxiv.org/abs/1309.6920}{{\ttfamily 1309.6920}}].

\bibitem{CDF:2009jgo}
{\scshape CDF} collaboration, \emph{{Evidence for a Narrow Near-Threshold
  Structure in the $J/\psi \, \phi$ Mass Spectrum in $B^+\to J/\psi \phi K^+$
  Decays}}, \href{https://doi.org/10.1103/PhysRevLett.102.242002}{\emph{Phys.\
  Rev.\ Lett.} {\bfseries {\bf 102}} (2009) 242002}
  [\href{https://arxiv.org/abs/0903.2229}{{\ttfamily 0903.2229}}].

\bibitem{LHCb:2016axx}
{\scshape LHCb} collaboration, \emph{{Observation of $J/\psi \, \phi$
  Structures Consistent with Exotic States from Amplitude Analysis of $B^+\to
  J/\psi \phi K^+$ Decays}},
  \href{https://doi.org/10.1103/PhysRevLett.118.022003}{\emph{Phys.\ Rev.\
  Lett.} {\bfseries {\bf 118}} (2017) 022003}
  [\href{https://arxiv.org/abs/1606.07895}{{\ttfamily 1606.07895}}].

\bibitem{LHCb:2021uow}
{\scshape LHCb} collaboration, \emph{{Observation of New Resonances Decaying to
  $J/\psi K^+$ and $J/\psi \, \phi$}},
  \href{https://doi.org/10.1103/PhysRevLett.127.082001}{\emph{Phys.\ Rev.\
  Lett.} {\bfseries {\bf 127}} (2021) 082001}
  [\href{https://arxiv.org/abs/2103.01803}{{\ttfamily 2103.01803}}].

\bibitem{LHCb:2016nsl}
{\scshape LHCb} collaboration, \emph{{Amplitude Analysis of $B^+\to J/\psi \phi
  K^+$ Decays}}, \href{https://doi.org/10.1103/PhysRevD.95.012002}{\emph{Phys.\
  Rev.\ D} {\bfseries {\bf 95}} (2017) 012002}
  [\href{https://arxiv.org/abs/1606.07898}{{\ttfamily 1606.07898}}].

\bibitem{CMS:2020eiw}
{\scshape CMS} collaboration, \emph{{Observation of the B$^0_{s}\to
  X(3872)\phi$ Decay}},
  \href{https://doi.org/10.1103/PhysRevLett.125.152001}{\emph{Phys.\ Rev.\
  Lett.} {\bfseries {\bf 125}} (2020) 152001}
  [\href{https://arxiv.org/abs/2005.04764}{{\ttfamily 2005.04764}}].

\bibitem{CMS:2019mny}
{\scshape CMS} collaboration, \emph{{Observation of the $\Lambda_{b}^0 \to$
  J/$\psi \Lambda \phi$ Decay in Proton-Proton Collisions at $\sqrt{s}=$
  13~TeV}}, \href{https://doi.org/10.1016/j.physletb.2020.135203}{\emph{Phys.\
  Lett.\ B} {\bfseries {\bf 802}} (2020) 135203}
  [\href{https://arxiv.org/abs/1911.03789}{{\ttfamily 1911.03789}}].

\bibitem{ATLAS:2014lga}
{\scshape ATLAS} collaboration, \emph{{Observation of an Excited $B_c^\pm$
  Meson State with the ATLAS Detector}},
  \href{https://doi.org/10.1103/PhysRevLett.113.212004}{\emph{Phys.\ Rev.\
  Lett.} {\bfseries {\bf 113}} (2014) 212004}
  [\href{https://arxiv.org/abs/1407.1032}{{\ttfamily 1407.1032}}].

\bibitem{CMS:2019uhm}
{\scshape CMS} collaboration, \emph{{Observation of Two Excited B$^+_{c}$
  States and Measurement of the B$^+_{c}(2S)$ Mass in pp Collisions at
  $\sqrt{s} =$ 13~TeV}},
  \href{https://doi.org/10.1103/PhysRevLett.122.132001}{\emph{Phys.\ Rev.\
  Lett.} {\bfseries {\bf 122}} (2019) 132001}
  [\href{https://arxiv.org/abs/1902.00571}{{\ttfamily 1902.00571}}].

\bibitem{LHCb:2019bem}
{\scshape LHCb} collaboration, \emph{{Observation of an Excited $B_c^+$
  State}}, \href{https://doi.org/10.1103/PhysRevLett.122.232001}{\emph{Phys.\
  Rev.\ Lett.} {\bfseries {\bf 122}} (2019) 232001}
  [\href{https://arxiv.org/abs/1904.00081}{{\ttfamily 1904.00081}}].

\bibitem{CMS:2020rcj}
{\scshape CMS} collaboration, \emph{{Measurement of B$_{c}(2S)^+$ and
  B$_{c}^*(2S)^+$ Cross Section Ratios in Proton-Proton Collisions at $\sqrt{s}
  =$ 13~TeV}}, \href{https://doi.org/10.1103/PhysRevD.102.092007}{\emph{Phys.\
  Rev.\ D} {\bfseries {\bf 102}} (2020) 092007}
  [\href{https://arxiv.org/abs/2008.08629}{{\ttfamily 2008.08629}}].

\bibitem{CMS:2021znk}
{\scshape CMS} collaboration, \emph{{Evidence for X$(3872)$ in Pb-Pb Collisions
  and Studies of its Prompt Production at $\sqrt {s_{NN}}$=5.02\,\,TeV}},
  \href{https://doi.org/10.1103/PhysRevLett.128.032001}{\emph{Phys.\ Rev.\
  Lett.} {\bfseries {\bf 128}} (2022) 032001}
  [\href{https://arxiv.org/abs/2102.13048}{{\ttfamily 2102.13048}}].

\bibitem{ATLAS:2022hsp}
{\scshape ATLAS} collaboration, \emph{{Snowmass White Paper Contribution:
  Physics with the Phase-2 ATLAS and CMS Detectors}},  in \emph{{2022 Snowmass
  Summer Study}}, 2022,
  \href{https://atlas.web.cern.ch/Atlas/GROUPS/PHYSICS/PUBNOTES/\\ATL-PHYS-PUB-2022-018/ATL-PHYS-PUB-2022-018.pdf}{https://atlas.web.cern.ch/Atlas/GROUPS/PHYSICS/PUBNOTES/\\ATL-PHYS-PUB-2022-018/ATL-PHYS-PUB-2022-018.pdf}.

\bibitem{ALICE:2022sli}
{\scshape ALICE} collaboration, \emph{{Measurement of the J/$\psi$ Polarization
  with Respect to the Event Plane in Pb-Pb Collisions at the LHC}},
  \href{https://arxiv.org/abs/2204.10171}{{\ttfamily 2204.10171}}.

\bibitem{ALICETPC:2020ann}
{\scshape ALICE TPC} collaboration, \emph{{The Upgrade of the ALICE TPC with
  GEMs and Continuous Readout}},
  \href{https://doi.org/10.1088/1748-0221/16/03/P03022}{\emph{JINST} {\bfseries
  {\bf 16}} (2021) P03022} [\href{https://arxiv.org/abs/2012.09518}{{\ttfamily
  2012.09518}}].

\bibitem{Adamova:2019vkf}
D.~Adamov\'a et~al., \emph{{A Next-Generation LHC Heavy-Ion Experiment}},
  \href{https://arxiv.org/abs/1902.01211}{{\ttfamily 1902.01211}}.

\bibitem{Colella:2022lmb}
{\scshape ALICE} collaboration, \emph{{Upgrade of the ALICE Experiment Beyond
  LHC Run 3}},  in \emph{{10th International Conference on New Frontiers in
  Physics}}, 2022 [\href{https://arxiv.org/abs/2206.13247}{{\ttfamily
  2206.13247}}].

\bibitem{ALICE3}
{\scshape ALICE} collaboration, ``{\em ALICE 3 Workshop}.''
  \url{https://indico.cern.ch/event/1063724/timetable/}, 2021.

\bibitem{BaBar:2005hhc}
{\scshape BaBar} collaboration, \emph{{Observation of a Broad Structure in the
  $\pi^+ \pi^- J/\psi$ Mass Spectrum around 4.26~GeV/c$^2$}},
  \href{https://doi.org/10.1103/PhysRevLett.95.142001}{\emph{Phys.\ Rev.\
  Lett.} {\bfseries {\bf 95}} (2005) 142001}
  [\href{https://arxiv.org/abs/hep-ex/0506081}{{\ttfamily hep-ex/0506081}}].

\bibitem{Belle:2007hrb}
{\scshape Belle} collaboration, \emph{{Observation of a Resonance-Like
  Structure in the $\pi^\pm \psi^\prime$ Mass Distribution in Exclusive $B \to
  K \pi^\pm \psi^\prime$ Decays}},
  \href{https://doi.org/10.1103/PhysRevLett.100.142001}{\emph{Phys.\ Rev.\
  Lett.} {\bfseries {\bf 100}} (2008) 142001}
  [\href{https://arxiv.org/abs/0708.1790}{{\ttfamily 0708.1790}}].

\bibitem{Olsen:2017bmm}
S.~Olsen, T.~Skwarnicki and D.~Zieminska, \emph{{Nonstandard Heavy Mesons and
  Baryons: Experimental Evidence}},
  \href{https://doi.org/10.1103/RevModPhys.90.015003}{\emph{Rev.\ Mod.\ Phys.}
  {\bfseries {\bf 90}} (2018) 015003}
  [\href{https://arxiv.org/abs/1708.04012}{{\ttfamily 1708.04012}}].

\bibitem{Belle:2011aa}
{\scshape Belle} collaboration, \emph{{Observation of Two Charged
  Bottomonium-Like Resonances in $\Upsilon$(5S) Decays}},
  \href{https://doi.org/10.1103/PhysRevLett.108.122001}{\emph{Phys.\ Rev.\
  Lett.} {\bfseries {\bf 108}} (2012) 122001}
  [\href{https://arxiv.org/abs/1110.2251}{{\ttfamily 1110.2251}}].

\bibitem{Belle:2019cbt}
{\scshape Belle} collaboration, \emph{{Observation of a New Structure Near
  10.75 GeV in the Energy Dependence of the $e^+e^-\to\Upsilon\pi^+\pi^-$
  (n=1,2,3) Cross Sections}},
  \href{https://doi.org/10.1007/JHEP10(2019)220}{\emph{JHEP} {\bfseries {\bf
  10}} (2019) 220} [\href{https://arxiv.org/abs/1905.05521}{{\ttfamily
  1905.05521}}].

\bibitem{Belle:2005lik}
{\scshape Belle} collaboration, \emph{{Observation of a New Charmonium State in
  Double Charmonium Production in $e^+e^-$ Annihilation at $\sqrt{s}~10.6$
  GeV}}, \href{https://doi.org/10.1103/PhysRevLett.98.082001}{\emph{Phys.\
  Rev.\ Lett.} {\bfseries {\bf 98}} (2007) 082001}
  [\href{https://arxiv.org/abs/hep-ex/0507019}{{\ttfamily hep-ex/0507019}}].

\bibitem{Belle:2007woe}
{\scshape Belle} collaboration, \emph{{Production of New Charmoniumlike States
  $e^+e^-\to J/\psi D^{(*)}\bar{D}^{(*)}$ at $\sqrt{s} \approx 10$ GeV}},
  \href{https://doi.org/10.1103/PhysRevLett.100.202001}{\emph{Phys.\ Rev.\
  Lett.} {\bfseries {\bf 100}} (2008) 202001}
  [\href{https://arxiv.org/abs/0708.3812}{{\ttfamily 0708.3812}}].

\bibitem{Belle:2007umv}
{\scshape Belle} collaboration, \emph{{Observation of Two Resonant Structures
  in $e^+e^-\to\pi^+ \pi^- \psi(2S)$ via Initial State Radiation at Belle}},
  \href{https://doi.org/10.1103/PhysRevLett.99.142002}{\emph{Phys.\ Rev.\
  Lett.} {\bfseries {\bf 99}} (2007) 142002}
  [\href{https://arxiv.org/abs/0707.3699}{{\ttfamily 0707.3699}}].

\bibitem{Belle:2013yex}
{\scshape Belle} collaboration, \emph{{Study of $e^+e^-\to\pi^+ \pi^- J/\psi$
  and Observation of a Charged Charmoniumlike State at Belle}},
  \href{https://doi.org/10.1103/PhysRevLett.110.252002}{\emph{Phys.\ Rev.\
  Lett.} {\bfseries {\bf 110}} (2013) 252002}
  [\href{https://arxiv.org/abs/1304.0121}{{\ttfamily 1304.0121}}].

\bibitem{BESIII:2013ris}
{\scshape BESIII} collaboration, \emph{{Observation of a Charged Charmoniumlike
  Structure in $e^+e^- \to \pi^+ \pi^-$ J/\ensuremath{\psi} at $\sqrt{s}$ =4.26
  GeV}}, \href{https://doi.org/10.1103/PhysRevLett.110.252001}{\emph{Phys. Rev.
  Lett.} {\bfseries 110} (2013) 252001}
  [\href{https://arxiv.org/abs/1303.5949}{{\ttfamily 1303.5949}}].

\bibitem{Belle:2015tbu}
{\scshape Belle} collaboration, \emph{{Energy Scan of the $e^+e^- \to
  h_b(nP)\pi^+\pi^-$ $(n=1,2)$ Cross Sections and Evidence for
  $\Upsilon(11020)$ Decays into Charged Bottomonium-Like States}},
  \href{https://doi.org/10.1103/PhysRevLett.117.142001}{\emph{Phys.\ Rev.\
  Lett.} {\bfseries {\bf 117}} (2016) 142001}
  [\href{https://arxiv.org/abs/1508.06562}{{\ttfamily 1508.06562}}].

\bibitem{BelleIIwhitepaper}
{\scshape Belle II} collaboration, \emph{{Belle II Physics Reach and Plans for
  the Next Decade and \\Beyond}},  in \emph{{2022 Snowmass Summer Study}}, 2022
  [\href{https://arxiv.org/abs/2207.06307}{{\ttfamily 2207.06307}}].

\bibitem{Asner:2022axe}
D.~Asner et~al., \emph{{Belle II Executive Summary}},  in \emph{{2022 Snowmass
  Summer Study}}, 2022 [\href{https://arxiv.org/abs/2203.10203}{{\ttfamily
  2203.10203}}].

\bibitem{Forti:2022mti}
{\scshape Belle II} collaboration, \emph{{Snowmass Whitepaper: The Belle II
  Detector Upgrade Program}},  in \emph{{2022 Snowmass Summer Study}}, 2022
  [\href{https://arxiv.org/abs/2203.11349}{{\ttfamily 2203.11349}}].

\bibitem{Natochii:2022vcs}
A.~Natochii et~al., \emph{{Beam Background Expectations for Belle II at
  SuperKEKB}},  in \emph{{2022 Snowmass Summer Study}}, 2022
  [\href{https://arxiv.org/abs/2203.05731}{{\ttfamily 2203.05731}}].

\bibitem{BESIII:2022mxl}
{\scshape BESIII} collaboration, \emph{{Physics in the Tau-Charm Region at
  BESIII}},  in \emph{{2022 Snowmass Summer Study}}, 2022
  [\href{https://arxiv.org/abs/2204.08943}{{\ttfamily 2204.08943}}].

\bibitem{BESIII:2016bnd}
{\scshape BESIII} collaboration, \emph{{Precise Measurement of the $e^+e^-\to
  \pi^+\pi^-J/\psi$ Cross Section at Center-of-Mass Energies from 3.77 to 4.60
  GeV}}, \href{https://doi.org/10.1103/PhysRevLett.118.092001}{\emph{Phys.\
  Rev.\ Lett.} {\bfseries {\bf 118}} (2017) 092001}
  [\href{https://arxiv.org/abs/1611.01317}{{\ttfamily 1611.01317}}].

\bibitem{BESIII:2016adj}
{\scshape BESIII} collaboration, \emph{{Evidence of Two Resonant Structures in
  $e^+ e^- \to \pi^+ \pi^- h_c$}},
  \href{https://doi.org/10.1103/PhysRevLett.118.092002}{\emph{Phys.\ Rev.\
  Lett.} {\bfseries {\bf 118}} (2017) 092002}
  [\href{https://arxiv.org/abs/1610.07044}{{\ttfamily 1610.07044}}].

\bibitem{BESIII:2014rja}
{\scshape BESIII} collaboration, \emph{{Study of $e^+e^-\to\omega\chi_{cJ}$ at
  Center-of-Mass Energies from 4.21 to 4.42 GeV}},
  \href{https://doi.org/10.1103/PhysRevLett.114.092003}{\emph{Phys.\ Rev.\
  Lett.} {\bfseries {\bf 114}} (2015) 092003}
  [\href{https://arxiv.org/abs/1410.6538}{{\ttfamily 1410.6538}}].

\bibitem{BESIII:2018iea}
{\scshape BESIII} collaboration, \emph{{Evidence of a Resonant Structure in the
  $e^+e^-\to \pi^+D^0D^{*-}$ Cross Section between 4.05 and 4.60 GeV}},
  \href{https://doi.org/10.1103/PhysRevLett.122.102002}{\emph{Phys.\ Rev.\
  Lett.} {\bfseries {\bf 122}} (2019) 102002}
  [\href{https://arxiv.org/abs/1808.02847}{{\ttfamily 1808.02847}}].

\bibitem{BESIII:2017tqk}
{\scshape BESIII} collaboration, \emph{{Measurement of $e^{+}e^{-}\rightarrow
  \pi^{+}\pi^{-}\psi(3686)$ from 4.008 to 4.600~GeV and Observation of a
  Charged Structure in the $\pi^{\pm}\psi(3686)$ Mass Spectrum}},
  \href{https://doi.org/10.1103/PhysRevD.96.032004}{\emph{Phys.\ Rev.\ D}
  {\bfseries {\bf 96}} (2017) 032004}
  [\href{https://arxiv.org/abs/1703.08787}{{\ttfamily 1703.08787}}].

\bibitem{BESIII:2021njb}
{\scshape BESIII} collaboration, \emph{{Cross Section Measurement of
  $e^+e^-\rightarrow\pi^+\pi^-(3686)$ from $\sqrt{s}$=4.0076 to 4.6984~GeV}},
  \href{https://doi.org/10.1103/PhysRevD.104.052012}{\emph{Phys.\ Rev.\ D}
  {\bfseries {\bf 104}} (2021) 052012}
  [\href{https://arxiv.org/abs/2107.09210}{{\ttfamily 2107.09210}}].

\bibitem{BESIII:2013ouc}
{\scshape BESIII} collaboration, \emph{{Observation of a Charged Charmoniumlike
  Structure $Z_c$(4020) and Search for the $Z_c$(3900) in $e^+e^- \to \pi^+
  \pi^- h_c$}},
  \href{https://doi.org/10.1103/PhysRevLett.111.242001}{\emph{Phys.\ Rev.\
  Lett.} {\bfseries {\bf 111}} (2013) 242001}
  [\href{https://arxiv.org/abs/1309.1896}{{\ttfamily 1309.1896}}].

\bibitem{BESIII:2013qmu}
{\scshape BESIII} collaboration, \emph{{Observation of a Charged
  $(D\bar{D}^{*})^\pm$ Mass Peak in $e^{+}e^{-} \to \pi D\bar{D}^{*}$ at
  $\sqrt{s} =$ 4.26 GeV}},
  \href{https://doi.org/10.1103/PhysRevLett.112.022001}{\emph{Phys.\ Rev.\
  Lett.} {\bfseries {\bf 112}} (2014) 022001}
  [\href{https://arxiv.org/abs/1310.1163}{{\ttfamily 1310.1163}}].

\bibitem{BESIII:2013mhi}
{\scshape BESIII} collaboration, \emph{{Observation of a Charged Charmoniumlike
  Structure in $e^+e^- \to (D^{*} \bar{D}^{*})^{\pm} \pi^\mp$ at
  $\sqrt{s}=$4.26~GeV}},
  \href{https://doi.org/10.1103/PhysRevLett.112.132001}{\emph{Phys.\ Rev.\
  Lett.} {\bfseries {\bf 112}} (2014) 132001}
  [\href{https://arxiv.org/abs/1308.2760}{{\ttfamily 1308.2760}}].

\bibitem{BESIII:2014gnk}
{\scshape BESIII} collaboration, \emph{{Observation of $e^+e^- \to \pi^0 \pi^0
  h_c$ and a Neutral Charmoniumlike Structure $Z_c($4020$)^0$}},
  \href{https://doi.org/10.1103/PhysRevLett.113.212002}{\emph{Phys.\ Rev.\
  Lett.} {\bfseries {\bf 113}} (2014) 212002}
  [\href{https://arxiv.org/abs/1409.6577}{{\ttfamily 1409.6577}}].

\bibitem{BESIII:2015cld}
{\scshape BESIII} collaboration, \emph{{Observation of $Z_c($3900$)^{0}$ in
  $e^+e^-\to\pi^0\pi^0 J/\psi$}},
  \href{https://doi.org/10.1103/PhysRevLett.115.112003}{\emph{Phys.\ Rev.\
  Lett.} {\bfseries {\bf 115}} (2015) 112003}
  [\href{https://arxiv.org/abs/1506.06018}{{\ttfamily 1506.06018}}].

\bibitem{BESIII:2020qkh}
{\scshape BESIII} collaboration, \emph{{Observation of a Near-Threshold
  Structure in the $K^+$ Recoil-Mass Spectra in $e^+e^- \rightarrow
  K^+(D_s^-D^{*0}+D_s^{*-}D^0$)}},
  \href{https://doi.org/10.1103/PhysRevLett.126.102001}{\emph{Phys.\ Rev.\
  Lett.} {\bfseries {\bf 126}} (2021) 102001}
  [\href{https://arxiv.org/abs/2011.07855}{{\ttfamily 2011.07855}}].

\bibitem{BESIII:2013fnz}
{\scshape BESIII} collaboration, \emph{{Observation of $e^+ e^- \to \gamma
  X$(3872) at BESIII}},
  \href{https://doi.org/10.1103/PhysRevLett.112.092001}{\emph{Phys.\ Rev.\
  Lett.} {\bfseries {\bf 112}} (2014) 092001}
  [\href{https://arxiv.org/abs/1310.4101}{{\ttfamily 1310.4101}}].

\bibitem{BESIII:2019esk}
{\scshape BESIII} collaboration, \emph{{Observation of the Decay $X(3872) \to
  \pi^0 \chi_{c1}(1P)$}},
  \href{https://doi.org/10.1103/PhysRevLett.122.202001}{\emph{Phys.\ Rev.\
  Lett.} {\bfseries {\bf 122}} (2019) 202001}
  [\href{https://arxiv.org/abs/1901.03992}{{\ttfamily 1901.03992}}].

\bibitem{BESIII:2019qvy}
{\scshape BESIII} collaboration, \emph{{Study of $e^+e^- \to \gamma \, \omega
  J/\psi$ and Observation of $X(3872) \to \omega J/\psi$}},
  \href{https://doi.org/10.1103/PhysRevLett.122.232002}{\emph{Phys.\ Rev.\
  Lett.} {\bfseries {\bf 122}} (2019) 232002}
  [\href{https://arxiv.org/abs/1903.04695}{{\ttfamily 1903.04695}}].

\bibitem{BESIII:2015iqd}
{\scshape BESIII} collaboration, \emph{{Observation of the $\psi(1^3D_2)$ State
  in $e^+e^-\to\pi^+\pi^-\gamma\chi_{c1}$ at BESIII}},
  \href{https://doi.org/10.1103/PhysRevLett.115.011803}{\emph{Phys.\ Rev.\
  Lett.} {\bfseries 115} (2015) 011803}
  [\href{https://arxiv.org/abs/1503.08203}{{\ttfamily 1503.08203}}].

\bibitem{BESIII:2022riz}
{\scshape BESIII} collaboration, \emph{{Observation of an Isoscalar Resonance
  with Exotic $J^{PC}=1^{-+}$ Quantum Numbers in
  $J/\psi\rightarrow\gamma\eta\eta'$}},
  \href{https://arxiv.org/abs/2202.00621}{{\ttfamily 2202.00621}}.

\bibitem{Luo:IPAC2018-MOPML013}
Q.~Luo and D.~Xu, \emph{{P}rogress on {P}reliminary {C}onceptual {S}tudy of
  {HIEPA}, a {S}uper {T}au{-C}harm {F}actory in {C}hina},  in \emph{Proc.\
  9$^{th}$ International Particle Accelerator Conference (IPAC '18), Vancouver,
  BC, Canada, April 29--May 4, 2018}, no.~9 in International Particle
  Accelerator Conference, (Geneva, Switzerland), p.~422, JACoW Publishing,
  2018,
  \href{{https://doi.org/10.18429/JACoW-IPAC2018-MOPML013}}{{https://doi.org/10.18429/JACoW-IPAC2018-MOPML013}}.

\bibitem{Barnyakov:2020vob}
{\scshape Super Charm-Tau Factory} collaboration, \emph{{The Project of the
  Super Charm-Tau Factory in Novosibirsk}},
  \href{https://doi.org/10.1088/1742-6596/1561/1/012004}{\emph{J. Phys.\ Conf.\
  Ser.} {\bfseries {\bf 1561}} (2020) 012004}.

\bibitem{Guo:2022kdi}
F.-K.~Guo, H.~Peng, J.-J.~Xie and X.~Zhou, \emph{{Hadron Spectroscopy at
  STCF}},  in \emph{{2022 Snowmass Summer Study}}, 2022
  [\href{https://arxiv.org/abs/2203.07141}{{\ttfamily 2203.07141}}].

\bibitem{Albaladejo:2022dwv}
M.~Albaladejo, L.~Bibrzycki, S.~Dobbs, C.~Fern\'andez-Ram\'\i{}rez,
  A.~Hiller~Blin, V.~Mathieu et~al., \emph{{Hadron Spectroscopy in
  Photoproduction}},  in \emph{{2022 Snowmass Summer Study}}, 2022
  [\href{https://arxiv.org/abs/2203.08290}{{\ttfamily 2203.08290}}].

\bibitem{GlueX:2019mkq}
{\scshape GlueX} collaboration, \emph{{First Measurement of Near-Threshold
  J/\ensuremath{\psi} Exclusive Photoproduction off the Proton}},
  \href{https://doi.org/10.1103/PhysRevLett.123.072001}{\emph{Phys.\ Rev.\
  Lett.} {\bfseries {\bf 123}} (2019) 072001}
  [\href{https://arxiv.org/abs/1905.10811}{{\ttfamily 1905.10811}}].

\bibitem{GlueX:2020idb}
{\scshape GlueX} collaboration, \emph{{The GlueX Beamline and Detector}},
  \href{https://doi.org/10.1016/j.nima.2020.164807}{\emph{Nucl.\ Instrum.\
  Meth.\ A} {\bfseries {\bf 987}} (2021) 164807}
  [\href{https://arxiv.org/abs/2005.14272}{{\ttfamily 2005.14272}}].

\bibitem{Hamdi:2019dbr}
{\scshape GlueX} collaboration, \emph{{Search for Exotic States in
  Photoproduction at GlueX}},
  \href{https://doi.org/10.1088/1742-6596/1667/1/012012}{\emph{J. Phys.\ Conf.\
  Ser.} {\bfseries {\bf 1667}} (2020) 012012}
  [\href{https://arxiv.org/abs/1908.11786}{{\ttfamily 1908.11786}}].

\bibitem{Arrington:2021alx}
J.~Arrington et~al., \emph{{Physics with CEBAF at 12~GeV and Future
  Opportunities}},  \href{https://arxiv.org/abs/2112.00060}{{\ttfamily
  2112.00060}}.

\bibitem{AbdulKhalek:2021gbh}
R.~Abdul~Khalek et~al., \emph{{Science Requirements and Detector Concepts for
  the Electron-Ion Collider: EIC Yellow Report}},
  \href{https://arxiv.org/abs/2103.05419}{{\ttfamily 2103.05419}}.

\bibitem{Glazier:2020}
D.~Glazier et~al., ``{{\scriptsize EL}SPECTRO: An Event Generator Framework for
  Incorporating Spectroscopy into Electro/Photoproduction Reactions}.''
  \url{https://github.com/dglazier/elSpectro}, 2020.

\bibitem{PANDA:2021ozp}
{\scshape PANDA} collaboration, \emph{{PANDA Phase One}},
  \href{https://doi.org/10.1140/epja/s10050-021-00475-y}{\emph{Eur.\ Phys.\ J.
  A} {\bfseries {\bf 57}} (2021) 184}
  [\href{https://arxiv.org/abs/2101.11877}{{\ttfamily 2101.11877}}].

\bibitem{Voloshin:1976ap}
M.~Voloshin and L.~Okun, \emph{{Hadron Molecules and Charmonium Atom}},
  {\emph{JETP Lett.} {\bfseries {\bf 23}} (1976)
  333,~\url{http://jetpletters.ru/ps/1801/article\_27526.shtml}}.

\bibitem{DeRujula:1976zlg}
A.~De~Rujula, H.~Georgi and S.~Glashow, \emph{{Molecular Charmonium: A New
  Spectroscopy?}},
  \href{https://doi.org/10.1103/PhysRevLett.38.317}{\emph{Phys.\ Rev.\ Lett.}
  {\bfseries {\bf 38}} (1977) 317}.

\bibitem{Jaffe:1976ig}
R.~Jaffe, \emph{{Multi-Quark Hadrons. 1. The Phenomenology of (2 Quark 2
  Anti-Quark) Mesons}},
  \href{https://doi.org/10.1103/PhysRevD.15.267}{\emph{Phys.\ Rev.\ D}
  {\bfseries {\bf 15}} (1977) 267}.

\bibitem{Jaffe:1976ih}
R.~Jaffe, \emph{{Multi-Quark Hadrons. 2. Methods}},
  \href{https://doi.org/10.1103/PhysRevD.15.281}{\emph{Phys.\ Rev.\ D}
  {\bfseries {\bf 15}} (1977) 281}.

\bibitem{Wilczek:2004im}
F.~Wilczek, \emph{{Diquarks as Inspiration and as Objects}},  in
  \emph{{Deserfest: A Celebration of the Life and Works of Stanley Deser}},
  p.~322, 2004,
  \href{{https://doi.org/10.1142/9789812775344\_0007}}{{https://doi.org/10.1142/9789812775344\_0007}}
  [\href{https://arxiv.org/abs/hep-ph/0409168}{{\ttfamily hep-ph/0409168}}].

\bibitem{Brambilla:2022ura}
M.~Karliner, E.~Santopinto et~al., \emph{{Substructure of Multiquark Hadrons
  (Snowmass 2021 White Paper)}},  in \emph{{2022 Snowmass Summer Study}}, 2022
  [\href{https://arxiv.org/abs/2203.16583}{{\ttfamily 2203.16583}}].

\bibitem{Chen:2018kuu}
H.-X.~Chen, C.-P.~Shen and S.-L.~Zhu, \emph{{A Possible Partner State of the
  $Y(2175)$}}, \href{https://doi.org/10.1103/PhysRevD.98.014011}{\emph{Phys.\
  Rev.\ D} {\bfseries {\bf 98}} (2018) 014011}
  [\href{https://arxiv.org/abs/1805.06100}{{\ttfamily 1805.06100}}].

\bibitem{BESIII:2021lho}
{\scshape BESIII} collaboration, \emph{{Measurement of Cross Section of $e^+
  e^- \to \phi \pi^+ \pi^-$ at Center-of-Mass Energies $\sqrt{s}$=2.0000-3.0800
  GeV}},  \href{https://arxiv.org/abs/2112.13219}{{\ttfamily 2112.13219}}.

\bibitem{Dubynskiy:2008mq}
S.~Dubynskiy and M.~Voloshin, \emph{{Hadro-Charmonium}},
  \href{https://doi.org/10.1016/j.physletb.2008.07.086}{\emph{Phys.\ Lett.\ B}
  {\bfseries {\bf 666}} (2008) 344}
  [\href{https://arxiv.org/abs/0803.2224}{{\ttfamily 0803.2224}}].

\bibitem{Dong:2020hxe}
X.-K.~Dong, F.-K.~Guo and B.-S.~Zou, \emph{{Explaining the Many Threshold
  Structures in the Heavy-Quark Hadron Spectrum}},
  \href{https://doi.org/10.1103/PhysRevLett.126.152001}{\emph{Phys.\ Rev.\
  Lett.} {\bfseries {\bf 126}} (2021) 152001}
  [\href{https://arxiv.org/abs/2011.14517}{{\ttfamily 2011.14517}}].

\bibitem{Swanson:2014tra}
E.~Swanson, \emph{{$Z_b$ and $Z_c$ Exotic States as Coupled Channel Cusps}},
  \href{https://doi.org/10.1103/PhysRevD.91.034009}{\emph{Phys.\ Rev.\ D}
  {\bfseries {\bf 91}} (2015) 034009}
  [\href{https://arxiv.org/abs/1409.3291}{{\ttfamily 1409.3291}}].

\bibitem{Pilloni:2016obd}
{\scshape JPAC} collaboration, \emph{{Amplitude Analysis and the Nature of the
  Z$_c$(3900)}},
  \href{https://doi.org/10.1016/j.physletb.2017.06.030}{\emph{Phys.\ Lett.\ B}
  {\bfseries {\bf 772}} (2017) 200}
  [\href{https://arxiv.org/abs/1612.06490}{{\ttfamily 1612.06490}}].

\bibitem{Guo:2019twa}
F.-K.~Guo, X.-H.~Liu and S.~Sakai, \emph{{Threshold Cusps and Triangle
  Singularities in Hadronic Reactions}},
  \href{https://doi.org/10.1016/j.ppnp.2020.103757}{\emph{Prog.\ Part.\ Nucl.\
  Phys.} {\bfseries {\bf 112}} (2020) 103757}
  [\href{https://arxiv.org/abs/1912.07030}{{\ttfamily 1912.07030}}].

\bibitem{Weinberg:1965zz}
S.~Weinberg, \emph{{Evidence That the Deuteron Is Not an Elementary Particle}},
  \href{https://doi.org/10.1103/PhysRev.137.B672}{\emph{Phys.\ Rev.} {\bfseries
  {\bf 137}} (1965) B672}.

\bibitem{Baru:2021ldu}
V.~Baru, X.-K.~Dong, M.-L.~Du, A.~Filin, F.-K.~Guo, C.~Hanhart et~al.,
  \emph{{Effective Range Expansion for Narrow Near-Threshold Resonances}},
  \href{https://arxiv.org/abs/2110.07484}{{\ttfamily 2110.07484}}.

\bibitem{Esposito:2021vhu}
A.~Esposito, L.~Maiani, A.~Pilloni, A.~Polosa and V.~Riquer, \emph{{From the
  Line Shape of the X(3872) to Its Structure}},
  \href{https://doi.org/10.1103/PhysRevD.105.L031503}{\emph{Phys.\ Rev.\ D}
  {\bfseries {\bf 105}} (2022) L031503}
  [\href{https://arxiv.org/abs/2108.11413}{{\ttfamily 2108.11413}}].

\bibitem{Takizawa:2012hy}
M.~Takizawa and S.~Takeuchi, \emph{{X(3872) As a Hybrid State of Charmonium and
  the Hadronic Molecule}},
  \href{https://doi.org/10.1093/ptep/ptt063}{\emph{PTEP} {\bfseries {\bf2013}}
  (2013) 093D01} [\href{https://arxiv.org/abs/1206.4877}{{\ttfamily
  1206.4877}}].

\bibitem{Wu:2010jy}
J.-J.~Wu, R.~Molina, E.~Oset and B.-S.~Zou, \emph{{Prediction of Narrow $N^*$
  and $\Lambda^*$ Resonances with Hidden Charm above 4 GeV}},
  \href{https://doi.org/10.1103/PhysRevLett.105.232001}{\emph{Phys.\ Rev.\
  Lett.} {\bfseries {\bf 105}} (2010) 232001}
  [\href{https://arxiv.org/abs/1007.0573}{{\ttfamily 1007.0573}}].

\bibitem{Guo:2019fdo}
F.-K.~Guo, H.-J.~Jing, U.-G.~Mei\ss{}ner and S.~Sakai, \emph{{Isospin Breaking
  Decays as a Diagnosis of the Hadronic Molecular Structure of the
  $P_c(4457)$}}, \href{https://doi.org/10.1103/PhysRevD.99.091501}{\emph{Phys.\
  Rev.\ D} {\bfseries {\bf 99}} (2019) 091501}
  [\href{https://arxiv.org/abs/1903.11503}{{\ttfamily 1903.11503}}].

\bibitem{Lebed:2017min}
R.~Lebed, \emph{{Spectroscopy of Exotic Hadrons Formed from Dynamical
  Diquarks}}, \href{https://doi.org/10.1103/PhysRevD.96.116003}{\emph{Phys.\
  Rev.\ D} {\bfseries {\bf 96}} (2017) 116003}
  [\href{https://arxiv.org/abs/1709.06097}{{\ttfamily 1709.06097}}].

\bibitem{Ferretti:2020ewe}
J.~Ferretti and E.~Santopinto, \emph{{Hidden-Charm and Bottom Tetra- and
  Pentaquarks with Strangeness in the Hadro-Quarkonium and Compact Tetraquark
  Models}}, \href{https://doi.org/10.1007/JHEP04(2020)119}{\emph{JHEP}
  {\bfseries {\bf 04}} (2020) 119}
  [\href{https://arxiv.org/abs/2001.01067}{{\ttfamily 2001.01067}}].

\bibitem{Maiani:2021tri}
L.~Maiani, A.~Polosa and V.~Riquer, \emph{{The New Resonances Z$_{cs}$(3985)
  and Z$_{cs}$(4003) (Almost) Fill Two Tetraquark Nonets of Broken SU(3)$_f$}},
  \href{https://doi.org/10.1016/j.scib.2021.04.040}{\emph{Sci.\ Bull.}
  {\bfseries {\bf 66}} (2021) 1616}
  [\href{https://arxiv.org/abs/2103.08331}{{\ttfamily 2103.08331}}].

\bibitem{Cleven:2015era}
M.~Cleven, F.-K.~Guo, C.~Hanhart, Q.~Wang and Q.~Zhao, \emph{{Employing Spin
  Symmetry to Disentangle Different Models for the XYZ States}},
  \href{https://doi.org/10.1103/PhysRevD.92.014005}{\emph{Phys.\ Rev.\ D}
  {\bfseries {\bf 92}} (2015) 014005}
  [\href{https://arxiv.org/abs/1505.01771}{{\ttfamily 1505.01771}}].

\bibitem{Hosaka:2016ypm}
A.~Hosaka, T.~Hyodo, K.~Sudoh, Y.~Yamaguchi and S.~Yasui, \emph{{Heavy Hadrons
  in Nuclear Matter}},
  \href{https://doi.org/10.1016/j.ppnp.2017.04.003}{\emph{Prog.\ Part.\ Nucl.\
  Phys.} {\bfseries {\bf 96}} (2017) 88}
  [\href{https://arxiv.org/abs/1606.08685}{{\ttfamily 1606.08685}}].

\bibitem{Braaten:2014qka}
E.~Braaten, C.~Langmack and D.~Smith, \emph{{Born-Oppenheimer Approximation for
  the XYZ Mesons}},
  \href{https://doi.org/10.1103/PhysRevD.90.014044}{\emph{Phys.\ Rev.\ D}
  {\bfseries {\bf 90}} (2014) 014044}
  [\href{https://arxiv.org/abs/1402.0438}{{\ttfamily 1402.0438}}].

\bibitem{Brambilla:2017uyf}
N.~Brambilla, G.~Krein, J.~Tarr\'us~Castell\`a and A.~Vairo,
  \emph{{Born-Oppenheimer Approximation in an Effective Field Theory
  Language}}, \href{https://doi.org/10.1103/PhysRevD.97.016016}{\emph{Phys.\
  Rev.\ D} {\bfseries {\bf 97}} (2018) 016016}
  [\href{https://arxiv.org/abs/1707.09647}{{\ttfamily 1707.09647}}].

\bibitem{Giron:2019cfc}
J.~Giron, R.~Lebed and C.~Peterson, \emph{{The Dynamical Diquark Model: Fine
  Structure and Isospin}},
  \href{https://doi.org/10.1007/JHEP01(2020)124}{\emph{JHEP} {\bfseries {\bf
  01}} (2020) 124} [\href{https://arxiv.org/abs/1907.08546}{{\ttfamily
  1907.08546}}].

\bibitem{Bedolla:2019zwg}
M.~Bedolla, J.~Ferretti, C.~Roberts and E.~Santopinto, \emph{{Spectrum of
  Fully-Heavy Tetraquarks from a Diquark+Antidiquark Perspective}},
  \href{https://doi.org/10.1140/epjc/s10052-020-08579-3}{\emph{Eur.\ Phys.\ J.
  C} {\bfseries {\bf 80}} (2020) 1004}
  [\href{https://arxiv.org/abs/1911.00960}{{\ttfamily 1911.00960}}].

\bibitem{Karliner:2014gca}
M.~Karliner and J.~Rosner, \emph{{Baryons with Two Heavy Quarks: Masses,
  Production, Decays, and Detection}},
  \href{https://doi.org/10.1103/PhysRevD.90.094007}{\emph{Phys.\ Rev.\ D}
  {\bfseries {\bf 90}} (2014) 094007}
  [\href{https://arxiv.org/abs/1408.5877}{{\ttfamily 1408.5877}}].

\bibitem{Karliner:2017qjm}
M.~Karliner and J.~Rosner, \emph{{Discovery of Doubly-Charmed $\Xi_{cc}$ Baryon
  Implies a Stable ($b b \bar{u} \bar{d}$) Tetraquark}},
  \href{https://doi.org/10.1103/PhysRevLett.119.202001}{\emph{Phys.\ Rev.\
  Lett.} {\bfseries {\bf 119}} (2017) 202001}
  [\href{https://arxiv.org/abs/1707.07666}{{\ttfamily 1707.07666}}].

\bibitem{Ader:1981db}
J.~Ader, J.~Richard and P.~Taxil, \emph{{Do Narrow Heavy Multi-Quark States
  Exist?}}, \href{https://doi.org/10.1103/PhysRevD.25.2370}{\emph{Phys.\ Rev.\
  D} {\bfseries {\bf 25}} (1982) 2370}.

\bibitem{Janc:2004qn}
D.~Janc and M.~Rosina, \emph{{The $T_{cc} = DD^*$ Molecular State}},
  \href{https://doi.org/10.1007/s00601-004-0068-9}{\emph{Few Body Syst.}
  {\bfseries {\bf 35}} (2004) 175}
  [\href{https://arxiv.org/abs/hep-ph/0405208}{{\ttfamily hep-ph/0405208}}].

\bibitem{Eichten:2017ffp}
E.~Eichten and C.~Quigg, \emph{{Heavy-Quark Symmetry Implies Stable Heavy
  Tetraquark Mesons $Q_iQ_j \bar q_k \bar q_l$}},
  \href{https://doi.org/10.1103/PhysRevLett.119.202002}{\emph{Phys.\ Rev.\
  Lett.} {\bfseries {\bf 119}} (2017) 202002}
  [\href{https://arxiv.org/abs/1707.09575}{{\ttfamily 1707.09575}}].

\bibitem{Meng:2020knc}
Q.~Meng, E.~Hiyama, A.~Hosaka, M.~Oka, P.~Gubler, K.~Can et~al., \emph{{Stable
  Double-Heavy Tetraquarks: Spectrum and Structure}},
  \href{https://doi.org/10.1016/j.physletb.2021.136095}{\emph{Phys.\ Lett.}
  {\bfseries {\bf B814}} (2021) 136095}
  [\href{https://arxiv.org/abs/2009.14493}{{\ttfamily 2009.14493}}].

\bibitem{Meng:2021yjr}
Q.~Meng, M.~Harada, E.~Hiyama, A.~Hosaka and M.~Oka, \emph{{Doubly Heavy
  Tetraquark Resonant States}},
  \href{https://doi.org/10.1016/j.physletb.2021.136800}{\emph{Phys.\ Lett.\ B}
  {\bfseries {\bf 824}} (2022) 136800}
  [\href{https://arxiv.org/abs/2106.11868}{{\ttfamily 2106.11868}}].

\bibitem{JPAC:2022ipt}
{\scshape JPAC} collaboration, \emph{{Snowmass White Paper: Need for Amplitude
  Analysis in the Discovery of New Hadrons}},  in \emph{{2022 Snowmass Summer
  Study}}, 2022 [\href{https://arxiv.org/abs/2203.08208}{{\ttfamily
  2203.08208}}].

\bibitem{Rodas:2021tyb}
{\scshape JPAC} collaboration, \emph{{Scalar and Tensor Resonances in $J/\psi $
  Radiative Decays}},
  \href{https://doi.org/10.1140/epjc/s10052-022-10014-8}{\emph{Eur.\ Phys.\ J.
  C} {\bfseries {\bf 82}} (2022) 80}
  [\href{https://arxiv.org/abs/2110.00027}{{\ttfamily 2110.00027}}].

\bibitem{JPAC:2017dbi}
{\scshape JPAC, COMPASS} collaboration, \emph{{New Analysis of $\eta\pi$ Tensor
  Resonances Measured at the COMPASS Experiment}},
  \href{https://doi.org/10.1016/j.physletb.2018.01.017}{\emph{Phys.\ Lett.\ B}
  {\bfseries {\bf 779}} (2018) 464}
  [\href{https://arxiv.org/abs/1707.02848}{{\ttfamily 1707.02848}}].

\bibitem{JPAC:2018zyd}
{\scshape JPAC} collaboration, \emph{{Determination of the Pole Position of the
  Lightest Hybrid Meson Candidate}},
  \href{https://doi.org/10.1103/PhysRevLett.122.042002}{\emph{Phys.\ Rev.\
  Lett.} {\bfseries {\bf 122}} (2019) 042002}
  [\href{https://arxiv.org/abs/1810.04171}{{\ttfamily 1810.04171}}].

\bibitem{Fernandez-Ramirez:2019koa}
{\scshape JPAC} collaboration, \emph{{Interpretation of the LHCb
  $P_c$(4312)$^+$ Signal}},
  \href{https://doi.org/10.1103/PhysRevLett.123.092001}{\emph{Phys.\ Rev.\
  Lett.} {\bfseries {\bf 123}} (2019) 092001}
  [\href{https://arxiv.org/abs/1904.10021}{{\ttfamily 1904.10021}}].

\bibitem{Albaladejo:2015lob}
M.~Albaladejo, F.-K.~Guo, C.~Hidalgo-Duque and J.~Nieves, \emph{{$Z_c(3900)$:
  What Has Been Really Seen?}},
  \href{https://doi.org/10.1016/j.physletb.2016.02.025}{\emph{Phys.\ Lett.\ B}
  {\bfseries {\bf 755}} (2016) 337}
  [\href{https://arxiv.org/abs/1512.03638}{{\ttfamily 1512.03638}}].

\bibitem{Mai:2017vot}
M.~Mai, B.~Hu, M.~D{\"o}ring, A.~Pilloni and A.~Szczepaniak, \emph{{Three-body
  Unitarity with Isobars Revisited}},
  \href{https://doi.org/10.1140/epja/i2017-12368-4}{\emph{Eur.\ Phys.\ J. A}
  {\bfseries {\bf 53}} (2017) 177}
  [\href{https://arxiv.org/abs/1706.06118}{{\ttfamily 1706.06118}}].

\bibitem{Jackura:2018xnx}
{\scshape JPAC} collaboration, \emph{{Phenomenology of Relativistic $3 \to 3$
  Reaction Amplitudes within the Isobar Approximation}},
  \href{https://doi.org/10.1140/epjc/s10052-019-6566-1}{\emph{Eur.\ Phys.\ J.
  C} {\bfseries {\bf 79}} (2019) 56}
  [\href{https://arxiv.org/abs/1809.10523}{{\ttfamily 1809.10523}}].

\bibitem{Mikhasenko:2019vhk}
M.~Mikhasenko, Y.~Wunderlich, A.~Jackura, V.~Mathieu, A.~Pilloni, B.~Ketzer
  et~al., \emph{{Three-Body Scattering: Ladders and Resonances}},
  \href{https://doi.org/10.1007/JHEP08(2019)080}{\emph{JHEP} {\bfseries {\bf
  08}} (2019) 080} [\href{https://arxiv.org/abs/1904.11894}{{\ttfamily
  1904.11894}}].

\bibitem{JPAC:2020umo}
{\scshape JPAC} collaboration, \emph{{$\omega \rightarrow 3\pi $ and $\omega
  \pi ^{0}$ Transition Form Factor Revisited}},
  \href{https://doi.org/10.1140/epjc/s10052-020-08576-6}{\emph{Eur.\ Phys.\ J.
  C} {\bfseries {\bf 80}} (2020) 1107}
  [\href{https://arxiv.org/abs/2006.01058}{{\ttfamily 2006.01058}}].

\bibitem{Bulava:2022ovd}
J.~Bulava et~al., \emph{{Hadron Spectroscopy with Lattice QCD}},  in
  \emph{{2022 Snowmass Summer Study}}, 2022
  [\href{https://arxiv.org/abs/2203.03230}{{\ttfamily 2203.03230}}].

\bibitem{Luscher:1986pf}
M.~L{\"u}scher, \emph{{Volume Dependence of the Energy Spectrum in Massive
  Quantum Field Theories. 2. Scattering States}},
  \href{https://doi.org/10.1007/BF01211097}{\emph{Commun.\ Math.\ Phys.}
  {\bfseries {\bf 105}} (1986) 153}.

\bibitem{Luscher:1990ux}
M.~L{\"u}scher, \emph{{Two Particle States on a Torus and Their Relation to the
  Scattering Matrix}},
  \href{https://doi.org/10.1016/0550-3213(91)90366-6}{\emph{Nucl.\ Phys.\ B}
  {\bfseries {\bf 354}} (1991) 531}.

\bibitem{Briceno:2016mjc}
R.~Brice\~no, J.~Dudek, R.~Edwards and D.~Wilson, \emph{{Isoscalar $\pi\pi$
  Scattering and the $\sigma$ Meson Resonance from QCD}},
  \href{https://doi.org/10.1103/PhysRevLett.118.022002}{\emph{Phys.\ Rev.\
  Lett.} {\bfseries {\bf 118}} (2017) 022002}
  [\href{https://arxiv.org/abs/1607.05900}{{\ttfamily 1607.05900}}].

\bibitem{Guo:2018zss}
D.~Guo, A.~Alexandru, R.~Molina, M.~Mai and M.~D\"oring, \emph{{Extraction of
  Isoscalar $\pi\pi$ Phase-Shifts from Lattice QCD}},
  \href{https://doi.org/10.1103/PhysRevD.98.014507}{\emph{Phys.\ Rev.\ D}
  {\bfseries {\bf 98}} (2018) 014507}
  [\href{https://arxiv.org/abs/1803.02897}{{\ttfamily 1803.02897}}].

\bibitem{Brett:2018jqw}
R.~Brett, J.~Bulava, J.~Fallica, A.~Hanlon, B.~H\"orz and C.~Morningstar,
  \emph{{Determination of $S$- and $P$-Wave $I=1/2$ $K\pi$ Scattering
  Amplitudes in $N_{\mathrm{f}}=2+1$ Lattice QCD}},
  \href{https://doi.org/10.1016/j.nuclphysb.2018.05.008}{\emph{Nucl.\ Phys.\ B}
  {\bfseries {\bf 932}} (2018) 29}
  [\href{https://arxiv.org/abs/1802.03100}{{\ttfamily 1802.03100}}].

\bibitem{Rendon:2020rtw}
G.~Rendon, L.~Leskovec, S.~Meinel, J.~Negele, S.~Paul, M.~Petschlies et~al.,
  \emph{{$I=1/2$ $S$-Wave and $P$-Wave $K\pi$ Scattering and the $\kappa$ and
  $K^*$ Resonances from Lattice QCD}},
  \href{https://doi.org/10.1103/PhysRevD.102.114520}{\emph{Phys.\ Rev.\ D}
  {\bfseries {\bf 102}} (2020) 114520}
  [\href{https://arxiv.org/abs/2006.14035}{{\ttfamily 2006.14035}}].

\bibitem{Wilson:2019wfr}
D.~Wilson, R.~Brice\~no, J.~Dudek, R.~Edwards and C.~Thomas, \emph{{The
  Quark-Mass Dependence of Elastic $\pi K$ Scattering from QCD}},
  \href{https://doi.org/10.1103/PhysRevLett.123.042002}{\emph{Phys.\ Rev.\
  Lett.} {\bfseries {\bf 123}} (2019) 042002}
  [\href{https://arxiv.org/abs/1904.03188}{{\ttfamily 1904.03188}}].

\bibitem{Gayer:2021xzv}
{\scshape Hadron Spectrum} collaboration, \emph{{Isospin-1/2 D\ensuremath{\pi}
  Scattering and the Lightest $ {D}_0^{\ast } $ Resonance from Lattice QCD}},
  \href{https://doi.org/10.1007/JHEP07(2021)123}{\emph{JHEP} {\bfseries {\bf
  07}} (2021) 123} [\href{https://arxiv.org/abs/2102.04973}{{\ttfamily
  2102.04973}}].

\bibitem{Moir:2016srx}
G.~Moir, M.~Peardon, S.~Ryan, C.~Thomas and D.~Wilson, \emph{{Coupled-Channel
  $D\pi$, $D\eta$ and $D_{s}\bar{K}$ Scattering from Lattice QCD}},
  \href{https://doi.org/10.1007/JHEP10(2016)011}{\emph{JHEP} {\bfseries {\bf
  10}} (2016) 011} [\href{https://arxiv.org/abs/1607.07093}{{\ttfamily
  1607.07093}}].

\bibitem{Mohler:2012na}
D.~Mohler, S.~Prelovsek and R.~Woloshyn, \emph{{$D \pi$ Scattering and $D$
  Meson Resonances from Lattice QCD}},
  \href{https://doi.org/10.1103/PhysRevD.87.034501}{\emph{Phys.\ Rev.\ D}
  {\bfseries {\bf 87}} (2013) 034501}
  [\href{https://arxiv.org/abs/1208.4059}{{\ttfamily 1208.4059}}].

\bibitem{Dudek:2014qha}
{\scshape Hadron Spectrum} collaboration, \emph{{Resonances in Coupled $\pi K
  -\eta K$ Scattering from Quantum Chromodynamics}},
  \href{https://doi.org/10.1103/PhysRevLett.113.182001}{\emph{Phys.\ Rev.\
  Lett.} {\bfseries {\bf 113}} (2014) 182001}
  [\href{https://arxiv.org/abs/1406.4158}{{\ttfamily 1406.4158}}].

\bibitem{Briceno:2017qmb}
R.~Brice\~no, J.~Dudek, R.~Edwards and D.~Wilson, \emph{{Isoscalar $\pi\pi,
  K\overline{K}, \eta\eta$ Scattering and the $\sigma, f_0, f_2$ Mesons from
  QCD}}, \href{https://doi.org/10.1103/PhysRevD.97.054513}{\emph{Phys.\ Rev.\
  D} {\bfseries {\bf 97}} (2018) 054513}
  [\href{https://arxiv.org/abs/1708.06667}{{\ttfamily 1708.06667}}].

\bibitem{Woss:2019hse}
A.~Woss, C.~Thomas, J.~Dudek, R.~Edwards and D.~Wilson, \emph{{$b_1$ Resonance
  in Coupled $\pi\omega$, $\pi\phi$ Scattering from Lattice QCD}},
  \href{https://doi.org/10.1103/PhysRevD.100.054506}{\emph{Phys.\ Rev.\ D}
  {\bfseries {\bf 100}} (2019) 054506}
  [\href{https://arxiv.org/abs/1904.04136}{{\ttfamily 1904.04136}}].

\bibitem{Prelovsek:2020eiw}
S.~Prelovsek, S.~Collins, D.~Mohler, M.~Padmanath and S.~Piemonte,
  \emph{{Charmonium-Like Resonances with J$^{PC}$ = 0$^{++}$, 2$^{++}$ in
  Coupled $ D\overline{D} $, $ D_s\overline{D}_{s} $ Scattering on the
  Lattice}}, \href{https://doi.org/10.1007/JHEP06(2021)035}{\emph{JHEP}
  {\bfseries {\bf 06}} (2021) 035}
  [\href{https://arxiv.org/abs/2011.02542}{{\ttfamily 2011.02542}}].

\bibitem{Bicudo:2015vta}
P.~Bicudo, K.~Cichy, A.~Peters, B.~Wagenbach and M.~Wagner, \emph{{Evidence for
  the Existence of $u d \bar{b} \bar{b}$ and the Non-Existence of $s s \bar{b}
  \bar{b}$ and $c c \bar{b} \bar{b}$ Tetraquarks from Lattice QCD}},
  \href{https://doi.org/10.1103/PhysRevD.92.014507}{\emph{Phys.\ Rev.\ D}
  {\bfseries {\bf 92}} (2015) 014507}
  [\href{https://arxiv.org/abs/1505.00613}{{\ttfamily 1505.00613}}].

\bibitem{Francis:2016hui}
A.~Francis, R.~Hudspith, R.~Lewis and K.~Maltman, \emph{{Lattice Prediction for
  Deeply Bound Doubly Heavy Tetraquarks}},
  \href{https://doi.org/10.1103/PhysRevLett.118.142001}{\emph{Phys.\ Rev.\
  Lett.} {\bfseries {\bf 118}} (2017) 142001}
  [\href{https://arxiv.org/abs/1607.05214}{{\ttfamily 1607.05214}}].

\bibitem{Hudspith:2020tdf}
R.~Hudspith, B.~Colquhoun, A.~Francis, R.~Lewis and K.~Maltman, \emph{{A
  Lattice Investigation of Exotic Tetraquark Channels}},
  \href{https://doi.org/10.1103/PhysRevD.102.114506}{\emph{Phys.\ Rev.\ D}
  {\bfseries {\bf 102}} (2020) 114506}
  [\href{https://arxiv.org/abs/2006.14294}{{\ttfamily 2006.14294}}].

\bibitem{Leskovec:2019ioa}
L.~Leskovec, S.~Meinel, M.~Pflaumer and M.~Wagner, \emph{{Lattice QCD
  Investigation of a Doubly-Bottom $\bar{b} \bar{b} u d$ Tetraquark with
  Quantum Numbers $I(J^P) = 0(1^+)$}},
  \href{https://doi.org/10.1103/PhysRevD.100.014503}{\emph{Phys.\ Rev.\ D}
  {\bfseries {\bf 100}} (2019) 014503}
  [\href{https://arxiv.org/abs/1904.04197}{{\ttfamily 1904.04197}}].

\bibitem{Junnarkar:2018twb}
P.~Junnarkar, N.~Mathur and M.~Padmanath, \emph{{Study of Doubly Heavy
  Tetraquarks in Lattice QCD}},
  \href{https://doi.org/10.1103/PhysRevD.99.034507}{\emph{Phys.\ Rev.\ D}
  {\bfseries {\bf 99}} (2019) 034507}
  [\href{https://arxiv.org/abs/1810.12285}{{\ttfamily 1810.12285}}].

\bibitem{HadronSpectrum:2012gic}
{\scshape Hadron Spectrum} collaboration, \emph{{Excited and Exotic Charmonium
  Spectroscopy from Lattice QCD}},
  \href{https://doi.org/10.1007/JHEP07(2012)126}{\emph{JHEP} {\bfseries {\bf
  07}} (2012) 126} [\href{https://arxiv.org/abs/1204.5425}{{\ttfamily
  1204.5425}}].

\bibitem{Ryan:2020iog}
{\scshape Hadron Spectrum} collaboration, \emph{{Excited and Exotic Bottomonium
  Spectroscopy from Lattice QCD}},
  \href{https://doi.org/10.1007/JHEP02(2021)214}{\emph{JHEP} {\bfseries {\bf
  02}} (2021) 214} [\href{https://arxiv.org/abs/2008.02656}{{\ttfamily
  2008.02656}}].

\bibitem{Prelovsek:2013cra}
S.~Prelovsek and L.~Leskovec, \emph{{Evidence for X(3872) from $DD^*$
  Scattering on the Lattice}},
  \href{https://doi.org/10.1103/PhysRevLett.111.192001}{\emph{Phys.\ Rev.\
  Lett.} {\bfseries {\bf 111}} (2013) 192001}
  [\href{https://arxiv.org/abs/1307.5172}{{\ttfamily 1307.5172}}].

\bibitem{Padmanath:2022cvl}
M.~Padmanath and S.~Prelovsek, \emph{{Evidence for a Doubly Charm Tetraquark
  Pole in $DD^*$ Scattering on the Lattice}},
  \href{https://arxiv.org/abs/2202.10110}{{\ttfamily 2202.10110}}.

\bibitem{Green:2021qol}
J.~Green, A.~Hanlon, P.M.~Junnarkar and H.~Wittig, \emph{{Weakly Bound H
  Dibaryon from SU(3)-Flavor-Symmetric QCD}},
  \href{https://doi.org/10.1103/PhysRevLett.127.242003}{\emph{Phys.\ Rev.\
  Lett.} {\bfseries {\bf 127}} (2021) 242003}
  [\href{https://arxiv.org/abs/2103.01054}{{\ttfamily 2103.01054}}].

\bibitem{Mai:2021nul}
{\scshape GWQCD} collaboration, \emph{{Three-Body Dynamics of the $a_{\it
  1}$(1260) Resonance from Lattice QCD}},
  \href{https://doi.org/10.1103/PhysRevLett.127.222001}{\emph{Phys.\ Rev.\
  Lett.} {\bfseries {\bf 127}} (2021) 222001}
  [\href{https://arxiv.org/abs/2107.03973}{{\ttfamily 2107.03973}}].

\bibitem{Briceno:2015dca}
R.~Brice\~no, J.~Dudek, R.~Edwards, C.~Shultz, C.~Thomas and D.~Wilson,
  \emph{{The Resonant $\pi^+\gamma\to\pi^+\pi^0$ Amplitude from Quantum
  Chromodynamics}},
  \href{https://doi.org/10.1103/PhysRevLett.115.242001}{\emph{Phys.\ Rev.\
  Lett.} {\bfseries {\bf 115}} (2015) 242001}
  [\href{https://arxiv.org/abs/1507.06622}{{\ttfamily 1507.06622}}].

\bibitem{Alexandrou:2018jbt}
C.~Alexandrou, L.~Leskovec, S.~Meinel, J.~Negele, S.~Paul, M.~Petschlies
  et~al., \emph{{$\pi\gamma \to \pi\pi$ Transition and the $\rho$ Radiative
  Decay Width from Lattice QCD}},
  \href{https://doi.org/10.1103/PhysRevD.98.074502}{\emph{Phys.\ Rev.\ D}
  {\bfseries {\bf 98}} (2018) 074502}
  [\href{https://arxiv.org/abs/1807.08357}{{\ttfamily 1807.08357}}].

\end{thebibliography}\endgroup

\end{document}